# Gender, Unpaid Work, and Social Norms in Young Italian Families: Evidence from Couples Time Diaries

C. Monfardini[1] and E. Pisanelli[2]

[1] University of Bologna, Department of Economics

[2] University of Bergamo, Department of Economics

This version: April 15, 2026.
Preliminary, please do not quote without permission of the authors

**Abstract**

Why do large gender inequalities in everyday life persist even as women strengthen their attach-ment to paid work? Existing evidence shows that women continue to do more unpaid work than men, but much of that evidence is based on individual diaries, says little about how inequality is jointly organized within couples, and rarely links daily time allocation to directly measured gender attitudes. This paper addresses that gap using the TIMES Observatory, an original survey of 1,928 co-resident couples with at least one child younger than 11 in Emilia-Romagna or Campania. The data combine matched partner diaries for one weekday and one weekend day with rich socio-economic information and direct measures of gender norms. We document three main findings. First, women do substantially more unpaid work and spend more time with children, while men do more paid work and enjoy more leisure without children. Second, these asymmetries remain sizeable even among dual full-time couples, implying that stronger female labor-market attachment does not by itself equalize daily life. Third, more traditional gender attitudes—especially among men—are descriptively associated with lower male participation in childcare and domestic work and with wider gaps in discretionary leisure. The analysis is descriptive rather than causal, but it shows that gender inequality within couples is visible not only in the amount of work performed, but also in the distribution of time that is genuinely discretionary.

This research is funded by the European Union – Next Generation EU, under the GRINS project – Growing Resilient, Inclusive and Sustainable (GRINS PE00000018 – CUP E63C22002140007, CUP J33C22002910001). The views and opinions expressed are solely those of the authors and do not necessarily reflect those of the European Union; therefore, the European Union cannot be held responsible for them.

Special thanks to Felix Bergmann for the excellent data processing work. Thanks also to the members of the research group — Francesca Barigozzi, Pietro Biroli, Margherita Fort, Natalia Montinari, Roberto Nisticò, and Valeria Zurla — who made the TIMES Observatory on young families' time use possible.

# 1. Introduction

Gender inequality is not only a question of wages, employment, or promotion. It is also visible in the way families organize the twenty-four hours of the day. In households with children, the allocation of time to paid work, domestic work, childcare, and leisure shapes not only future earnings and career trajectories, but also current well-being, autonomy, and the quality of family life. The daily division of time is therefore itself an outcome of substantive interest: it reveals how constraints, responsibilities, and opportunities are distributed within households.

This paper studies a simple puzzle. Over recent decades, women have strengthened their labor-market attachment in most high-income countries, including Italy. Yet convergence in paid work has not been matched by equivalent convergence in unpaid work and discretionary leisure. In many households, women have moved closer to men in market work without men moving equally closer to women in domestic work and childcare. As a result, apparent progress in equality can coexist with persistent asymmetries in everyday well-being.

A large literature already documents gender differences in time allocation. Research in family economics has long treated time as central to household production, specialization, bargaining, and identity [7, 15, 10, 11, 2, 3]. Cross-country diary evidence shows that women continue to do more unpaid work and men more paid work, although the magnitude of the gap varies across institutional and cultural settings [1, 14, 9, 13]. Recent work also shows that gender inequality extends beyond total leisure to the way leisure is experienced, including whether it occurs in the presence of children [16]. At the same time, policy-oriented research has emphasized that time-use indicators are central to the measurement of women's empowerment and to the recognition, reduction, and redistribution of unpaid care work [12].

Italy is a useful setting in which to revisit this issue. Previous evidence shows that Italy combines relatively low female employment, uneven childcare provision, and persistent traditional gender norms, all of which help sustain a wide domestic specialization gap. Using Italian time-use data, Barigozzi et al. [6] show that women increased their labor-market attachment between 2002 and 2014, but unpaid work remained highly asymmetric. Other Italian evidence finds that inequality at home narrows only under specific conditions, such as when women become primary earners or when local gender systems are less traditional [17]. Yet the most recent nationally representative Italian time-use survey is now dated, and standard sources do not allow researchers to jointly observe both partners' daily behavior and directly measured attitudes.

This paper addresses three gaps in that literature. First, many time-use studies observe only one adult per household, making it difficult to examine how inequality is jointly organized within couples. Second, the role of norms is often inferred indirectly rather than measured directly. Third, existing work has devoted less attention to constrained versus unconstrained leisure, even though this distinction is central for understanding autonomy and control over daily life.

We address these gaps using the TIMES Observatory on young families, an original survey of 1,928 co-resident couples with at least one child younger than 11 in Emilia-Romagna or Campania. TIMES has two features that are especially valuable for the present question. First, both partners complete two twenty-four-hour diaries, one for a weekday and one for a weekend day, allowing direct within-couple comparisons. Second, the diaries are linked to rich socio-economic information and to direct measures of gender-role attitudes, parenting beliefs, and household responsibilities. The survey also includes indicators of mental load and family organization, extending the analysis beyond visible time inputs alone [5].

We make two contributions. The first is measurement. By using matched partner diaries, we observe the joint organization of time within the same household rather than inferring inequality from separate male and female averages. The second is substantive. We show that gender inequality within couples is especially visible not only in total workloads, but in the distribution of discretionary and less constrained time. This shifts attention from time-use differences in general to inequality in autonomy within daily life.

Our findings are straightforward. Women do substantially more unpaid work and spend more time with children, while men do more paid work and enjoy more leisure without children. These asymmetries remain sizeable even among dual full-time couples, implying that stronger female labor-market attachment is not sufficient on its own to equalize daily life. We also show that more traditional gender attitudes—especially among men—are descriptively associated with lower male participation in childcare and domestic work and with wider gaps in discretionary leisure. The analysis is deliberately descriptive rather than causal, but it identifies a set of couple-level indicators that speak directly to gender inequality in everyday well-being.

The rest of the paper proceeds as follows. Section 2 situates the paper in the literature and clarifies the conceptual framing. Section 3 describes the TIMES data and the measurement strategy. Section 4 presents the descriptive evidence on time allocation and gender attitudes and then reports the descriptive relationship between norms and time use. Section 5 discusses the implications for the measurement of gender inequality and family well-being. Section 6 concludes.

## 2. Conceptual background and related literature

Three strands of literature are most relevant for the present paper.

The first comes from the economics of the family. In Becker-style models, household members allocate time according to comparative advantage, which can generate specialization within the couple [7]. In bargaining models, the resulting allocation depends on relative resources, outside options, and intrahousehold decision-making power [10, 11]. Identity-based approaches add a further margin: time allocation may also reflect the desire to conform to gendered expectations about what men and women are supposed to do within families [2]. Taken together, these frameworks imply that unequal time allocations need not be interpreted only as efficiency outcomes; they may also reflect social norms, unequal power, or both.

The second strand is the empirical time-use literature. Harmonized evidence for Europe and other high-income countries consistently shows that women spend more time in unpaid work and childcare, while men spend more time in paid work [1, 14, 13]. Burda, Hamermesh, and Weil are especially important because they show that apparent parity in total work can conceal large differences in the composition of work between paid and unpaid activities [9]. For parents, leisure adds a further layer. Recent evidence for 13 European countries shows that mothers not only have less leisure overall, but also spend more of that leisure with children, whereas fathers spend a larger share in more discretionary forms [16]. A full account of inequality therefore requires going beyond total work and examining how non-work time is constrained.

The third strand concerns gender norms, care infrastructure, and well-being. Policy-oriented work argues that time-use indicators are central to the measurement of women's empowerment because unpaid care work shapes autonomy, labor-market opportunities, and daily quality of life [12]. For Italy, the evidence points to a persistent domestic asymmetry that narrows only partially when mothers work and more clearly in less traditional territorial contexts [17, 6]. Other work based on new digital diaries shows the value of measuring not just the quantity of parental time but also the quality of parent-child interactions [8]. Recent TIMES-based research further shows that the invisible organizational responsibilities of family life—the mental load—are themselves highly gendered [5].

This paper sits at the intersection of these strands. Relative to the existing literature, its comparative advantage is not causal identification but measurement. Matched partner diaries make the within-couple organization of time directly observable. Direct measures of attitudes make it possible to relate time allocation to gender norms rather than inferring them indirectly. And the distinction between leisure with and without children speaks directly to a quality-of-life perspective, because it separates nominal non-work time from time that is genuinely discretionary.

We therefore adopt a deliberately modest conceptual framing. We do not test specialization, bargaining, or gender ideology as competing causal explanations. Instead, we use couple-level diary data to document how gender inequality appears in the everyday allocation of time and how those inequalities are associated with respondents' attitudes toward gender roles. The central claim is descriptive but substantive: gender inequality within couples is expressed not only through differences in paid and unpaid work, but through unequal access to time that is flexible, continuous, and more fully one's own. Three strands of literature are most relevant for the present paper.

The first strand comes from the economics of the family. In Becker-style models, household members allocate time according to comparative advantage, potentially generating specialization within the couple [7]. In bargaining models, the resulting allocation depends on relative resources, outside options, and intrahousehold decision-making power [10, 11]. Identity-based approaches add an important refinement: time allocation may also reflect the desire to conform to gendered expectations about what men and women are supposed to do within families [2]. These frameworks imply that unequal time allocations need not be interpreted only as efficiency outcomes; they may also reflect social norms, unequal power, or both.

The second strand is the empirical time-use literature. Harmonized evidence for Europe and other high-income countries consistently shows that women spend more time in unpaid work and childcare, while men spend more time in paid work [1, 14, 13]. Burda, Hamermesh, and Weil's work on total work is especially important because it highlights that apparent gender equality in aggregate work time can conceal stark differences in composition between paid and unpaid work [9]. For parents, leisure adds another relevant layer. Recent evidence for 13 European countries shows that mothers not only have less leisure overall, but also spend more of that leisure with children, whereas fathers spend a larger share in more discretionary forms [16]. This suggests that a full account of inequality requires going beyond total work and measuring how non-work time is constrained.

The third strand concerns gender norms, care infrastructure, and well-being. Policy-oriented work has argued that time-use indicators are central for measuring women's empowerment because unpaid care work shapes autonomy, labor-market opportunities, and daily quality of life [12]. For Italy, the evidence points to a persistent domestic asymmetry that narrows only partially when mothers work and only more clearly in less traditional territorial contexts [17, 6]. Other research based on new digital diaries shows the value of measuring not just the quantity of parental time but also the quality of parent-child interactions [8]. Recent TIMES-based work also shows that invisible organizational responsibilities—the mental load of coordinating family life—are highly gendered and strongly related to within-couple inequalities [5]. Together, these contributions suggest that gender inequality in households should be understood as multidimensional: it concerns visible time, invisible organization, and the distribution of autonomy.

This paper sits at the intersection of these strands. Relative to the existing literature, its main comparative advantage is not causal identification but measurement. The use of matched partner diaries makes the within-couple distribution of time directly observable. The integration of time diaries, attitudes, and indicators of family organization allows a richer account of inequality than is usually possible in standard time-use surveys. And the focus on leisure with and without children speaks directly to a quality-of-life perspective, because it distinguishes between time that is nominally non-work and time that is genuinely discretionary.

The paper therefore advances the literature in the following ways. It uses a novel combination of indicators to show how gender inequality is reproduced through the daily organization of family life. The central claim is not that one single mechanism explains all observed gaps, but that a set of measurable, couple-level indicators—paid work, unpaid work, time with children, leisure without children, and mental-load-related responsibilities—jointly reveal an unequal distribution of well-being inside households.

Rather than attempting to identify a single mechanism behind these patterns, we adopt a more modest conceptual framing. We do not test specialization, bargaining, or gender ideology as competing causal explanations. Instead, we use couple-level diary data to document how gender inequality appears in the everyday allocation of time and how those inequalities are associated with respondents' attitudes toward gender roles.

This framing centers two analytically distinct but related dimensions of inequality. The first is the unequal allocation of domestic labor and leisure across women and men within couples. The second is inequality in discretionary time, understood as time that is less constrained by obligation, more continuous, and more available for self-directed use. This distinction matters because equal amounts of nominal leisure do not necessarily imply equal autonomy: leisure that is fragmented, contingent, or easily interrupted is qualitatively different from leisure that is secure and self-directed.

Our analysis is therefore descriptive in a strong sense. It documents the organization of daily time within couples and examines whether more traditional gender attitudes are associated with lower male participation in domestic work and wider gender gaps in discretionary time. This narrower framing better matches the design and clarifies the paper's central claim: gender inequality in couples is expressed not only through time-use gaps, but through unequal access to autonomous time.

## 3. Data and measurement

### 3.1. Why TIMES matters

The main empirical source is the TIMES Observatory, an original survey of co-resident couples with at least one child younger than 11 living in Emilia-Romagna or Campania. The survey collects matched information for both partners, making the couple—rather than the individual respondent—the relevant unit of observation. Each partner completes a rich socio-economic questionnaire and two twenty-four-hour time diaries, one for a weekday and one for a weekend day, through a web-based interface. Activities are recorded in ten-minute intervals and include primary and secondary activities, co-presence, and indicators of whether children are involved in the episode.

Several design features make TIMES particularly valuable for the present question. The survey explicitly targets couples with young children, a life-course stage in which time pressure, care needs, and gender specialization are especially salient. The questionnaires and diaries are collected from both partners independently but linked at the household level, which allows direct within-couple comparisons. The survey also collects attitudinal information on gender roles, maternal employment, father involvement, and household organization. In addition, the broader TIMES measurement strategy includes indicators of organizational responsibility and mental load, which strengthen the interpretation of time inequality as a multidimensional phenomenon rather than a simple difference in hours [5]. Key features of the underlying data collection protocol—including the two-stage design, quota-based recruitment across provinces and municipality size, diary validation rules, and the focus on families with children under 11—are described in detail in the TIMES endline report [4]. The report also documents that the analytical sample used for couple-level analysis contains 3,864 individuals, that is, 1,928 couples who completed both the questionnaire and the two-day diary module [4, pp. 14–17].

This design creates a dataset that is unusually informative for the study of gender inequality as a quality-of-life issue. Standard harmonized time-use surveys often observe only one household member, whereas TIMES makes it possible to compare what each partner does during the day, how responsibilities are perceived, and how these behaviors correlate with attitudes. That feature is central for a paper that treats within-household time allocation as a social indicator rather than merely an individual behavioral outcome.

### 3.2. Sample and key variables

The full sample includes 1,928 couples. Appendix Table 7 shows that women are on average younger than men, substantially less likely to be employed, and less likely to work full-time conditional on employment. Overall, 52% of women are employed compared with 95% of men. Among the employed, 60% of women and 80% of men work full-time. The sample is almost evenly split between Emilia-Romagna and Campania, which makes it possible to contrast two regions with markedly different labor-market contexts and gender-role environments. The endline report confirms the same broad picture: male employment rates are substantially higher than female rates in both regions, especially in Campania, and gender gaps in unpaid work are visibly larger in the South than in Emilia-Romagna [4, pp. 14–17, 25–30].

The paper studies several indicators of time allocation. *Paid work* includes market work and commuting. *Unpaid work* combines childcare and domestic work. We also separately analyze *childcare*, *housework*, *leisure*, *leisure with children*, *leisure without children*, and *total time with children*. Appendix B provides detailed definitions. All time variables are measured in minutes per day. For many outcomes, we consider women's time, men's time, and the within-couple gap defined as women's time minus men's time.

Two choices are worth emphasizing. First, we distinguish between total leisure and leisure without children. The latter is interpreted as a closer proxy for discretionary time and autonomy. Second, we treat time with children and unpaid work not only as labor inputs but also as indicators of how family well-being is organized and of which partner bears the everyday constraints of care.

The central attitudinal variable is an index of traditional gender norms constructed from survey items on family roles, maternal employment, whether a child suffers if the mother works, whether men should contribute at home, and related statements. Higher values indicate more conservative attitudes. In the descriptive section, we also report the individual components of these attitudes. The endline report shows that women are systematically less supportive of conservative gender norms than men, with sharper gaps in Campania [4, pp. 16–17, 58–61].

### 3.3. Empirical strategy

The empirical analysis has two parts. First, we present descriptive comparisons of men's and women's time use and attitudes in the full sample, by region, and for the subsample in which both partners work full-time. This first step addresses the central descriptive puzzle of the paper: whether the combination of two full-time jobs is enough to compress gender gaps in unpaid work and discretionary leisure. Second, we estimate a battery of ordinary least squares regressions linking time-use outcomes to the traditional-norm indices of women and men, separately. For each outcome, we estimate models without controls and with a common set of socio-economic controls that include age of the oldest child, age of the mother, age of the father and having a college degree or more, and we allow the association to differ in Campania through interaction terms.

These regressions are descriptive. They should be interpreted as conditional correlations, not as evidence that norms causally determine time allocation. The purpose is to document whether more traditional attitudes co-occur with more gender-specialized arrangements and whether the most systematic margins of adjustment concern men's unpaid work and leisure without children rather than women's paid work. From a social-indicators perspective, the goal is to identify a coherent set of observable indicators that reveal how gender inequality is organized inside households.

## 4. Results

### 4.1. Gender differences in time use

Table 1 reports the main descriptive results for the full sample. Three findings stand out.

First, the familiar asymmetry between paid and unpaid work remains large. On weekdays, women perform 207.9 minutes of paid work against 427.1 minutes for men, a gap of -219.2 minutes. By contrast, women perform 339.4 minutes of unpaid work against 128.8 minutes for men, a gap of 210.5 minutes. On weekends, the paid-work gap narrows but remains substantial, while the unpaid-work gap is still
137.6 minutes. Women therefore continue to bear most domestic work even as men continue to perform most market work.

Second, total work masks these asymmetries. On weekdays, women and men perform almost the same amount of total work—547.3 versus 555.9 minutes—and the difference is not statistically significant. But this apparent parity is entirely compositional: men do more market work, women far more unpaid work. On weekends, the asymmetry becomes visible even in total work, with women doing about 51 more minutes of total work per day.

Third, the main difference in non-work time is not total leisure but the kind of leisure each partner enjoys. Total weekday leisure is almost identical for women and men, but its composition differs sharply. Women spend significantly more leisure time with children, whereas men enjoy more leisure without children. On weekends, men also enjoy more total leisure overall. Leisure accompanied by childcare is not equivalent to fully discretionary leisure, so the relevant inequality concerns not only how much non-work time each partner has, but how constrained that time is.

The decomposition of unpaid work clarifies the source of the gap. Women devote around 175 minutes to childcare on weekdays compared with 88 minutes for men, and 164.8 versus 117.9 minutes on weekends. In housework, the gap is larger still: 164.0 versus 40.3 minutes on weekdays and 160.0 versus 68.9 minutes on weekends. Housework thus remains the most feminized and least rebalanced component of unpaid work.

Regional comparisons reinforce this interpretation. The same qualitative pattern appears in both Emilia-Romagna and Campania, but the asymmetry is stronger in Campania. In that region, the weekday unpaid-work gap exceeds 260 minutes and the weekend gap approaches 185 minutes (Table 3). The descriptive evidence is therefore consistent with the view that weaker female labor-market attachment and more traditional gender norms are associated with stronger within-family specialization.

To make the norms results easier to evaluate, we summarize the main estimates in a compact table in the main text and use the coefficient plots as supplementary detail. The table reports the key associations between traditional gender attitudes and (i) men's domestic participation, (ii) women's and men's leisure time, and (iii) the couple-level gap in leisure or discretionary time. Presenting these estimates directly allows readers to assess magnitudes alongside direction and statistical uncertainty rather than relying primarily on visual inspection across a long sequence of figures.

Table 1: Gender differences in time use

| Activity | Women | Men | Diff. | t-statistic | p-value |
|---|---|---|---|---|---|
| **Paid work** | | | | | |
| Weekday | 207.9 | 427.1 | -219.2 | -27.39 | 0.000 |
| Weekend | 64.4 | 151.4 | -87.0 | -12.75 | 0.000 |
| **Unpaid work** | | | | | |
| Weekday | 339.4 | 128.8 | 210.5 | 35.04 | 0.000 |
| Weekend | 325.1 | 187.4 | 137.6 | 21.77 | 0.000 |
| **Total work** | | | | | |
| Weekday | 547.3 | 555.9 | -8.6 | -1.14 | 0.254 |
| Weekend | 389.5 | 338.8 | 50.7 | 6.22 | 0.000 |
| **Leisure** | | | | | |
| Weekday | 815.5 | 815.7 | -0.2 | -0.03 | 0.974 |
| Weekend | 952.4 | 1002.6 | -50.2 | -6.24 | 0.000 |
| **Leisure with children** | | | | | |
| Weekday | 194.6 | 156.9 | 37.8 | 6.80 | 0.000 |
| Weekend | 305.0 | 288.7 | 16.4 | 2.34 | 0.019 |
| **Leisure without children** | | | | | |
| Weekday | 620.8 | 658.8 | -38.0 | -6.60 | 0.000 |
| Weekend | 647.3 | 713.9 | -66.6 | -10.57 | 0.000 |
| **Childcare** | | | | | |
| Weekday | 174.8 | 88.1 | 86.7 | 20.38 | 0.000 |
| Weekend | 164.8 | 117.9 | 46.9 | 9.35 | 0.000 |
| **Housework** | | | | | |
| Weekday | 164.0 | 40.3 | 123.7 | 32.75 | 0.000 |
| Weekend | 160.0 | 68.9 | 91.2 | 23.19 | 0.000 |
| **Care for other people** | | | | | |
| Weekday | 0.6 | 0.5 | 0.1 | 0.40 | 0.692 |
| Weekend | 0.3 | 0.7 | -0.4 | -1.59 | 0.111 |
| **Total time with children** | | | | | |
| Weekday | 479.1 | 278.2 | 200.9 | 24.99 | 0.000 |
| Weekend | 620.9 | 499.7 | 121.2 | 14.18 | 0.000 |
| **Engaged time with children** | | | | | |
| Weekday | 387.0 | 240.6 | 146.3 | 20.60 | 0.000 |
| Weekend | 526.9 | 441.5 | 85.4 | 10.37 | 0.000 |
| **Quality time with children** | | | | | |
| Weekday | 267.4 | 157.5 | 109.9 | 20.94 | 0.000 |
| Weekend | 324.8 | 247.8 | 76.9 | 12.39 | 0.000 |

*Notes:* The table reports average daily time devoted to each activity, measured in minutes. The difference is computed as women minus men. t-statistics and p-values come from tests of equality of means across the two groups.

Table 2: Gender differences in time use (Emilia-Romagna)

| Activity | Women | Men | Diff. | t-statistic | p-value |
|---|---|---|---|---|---|
| **Paid work** | | | | | |
| Weekday | 266.1 | 426.2 | -160.1 | -13.54 | 0.000 |
| Weekend | 66.1 | 114.0 | -47.9 | -5.17 | 0.000 |
| **Unpaid work** | | | | | |
| Weekday | 272.1 | 112.9 | 159.2 | 20.42 | 0.000 |
| Weekend | 278.7 | 189.7 | 89.1 | 10.47 | 0.000 |
| **Total work** | | | | | |
| Weekday | 538.2 | 539.1 | -0.9 | -0.08 | 0.938 |
| Weekend | 344.8 | 303.6 | 41.2 | 3.72 | 0.000 |
| **Leisure** | | | | | |
| Weekday | 838.0 | 840.2 | -2.1 | -0.20 | 0.845 |
| Weekend | 994.8 | 1038.5 | -43.7 | -3.88 | 0.000 |
| **Leisure with children** | | | | | |
| Weekday | 196.8 | 160.7 | 36.1 | 4.25 | 0.000 |
| Weekend | 317.2 | 296.2 | 21.0 | 2.04 | 0.041 |
| **Leisure without children** | | | | | |
| Weekday | 641.3 | 679.5 | -38.3 | -4.33 | 0.000 |
| Weekend | 677.5 | 742.3 | -64.7 | -6.98 | 0.000 |
| **Childcare** | | | | | |
| Weekday | 139.8 | 76.7 | 63.1 | 11.91 | 0.000 |
| Weekend | 134.2 | 115.3 | 18.9 | 2.88 | 0.004 |
| **Housework** | | | | | |
| Weekday | 131.6 | 36.1 | 95.5 | 19.19 | 0.000 |
| Weekend | 144.4 | 73.9 | 70.5 | 12.94 | 0.000 |
| **Care for other people** | | | | | |
| Weekday | 0.7 | 0.1 | 0.6 | 1.73 | 0.083 |
| Weekend | 0.1 | 0.5 | -0.3 | -1.46 | 0.145 |
| **Total time with children** | | | | | |
| Weekday | 425.4 | 271.4 | 154.0 | 13.87 | 0.000 |
| Weekend | 606.0 | 513.9 | 92.2 | 7.65 | 0.000 |
| **Engaged time with children** | | | | | |
| Weekday | 356.4 | 241.2 | 115.2 | 11.54 | 0.000 |
| Weekend | 524.6 | 463.6 | 61.0 | 5.14 | 0.000 |
| **Quality time with children** | | | | | |
| Weekday | 230.8 | 151.0 | 79.8 | 11.33 | 0.000 |
| Weekend | 309.8 | 250.9 | 58.9 | 6.81 | 0.000 |

*Notes:* The table reports average daily time devoted to each activity, measured in minutes. The difference is computed as women minus men. t-statistics and p-values come from tests of equality of means across the two groups.

Table 3: Gender differences in time use (Campania)

| Activity | Women | Men | Diff. | t-statistic | p-value |
|---|---|---|---|---|---|
| **Paid work** | | | | | |
| Weekday | 151.3 | 427.9 | -276.7 | -26.34 | 0.000 |
| Weekend | 62.8 | 187.8 | -125.1 | -12.67 | 0.000 |
| **Unpaid work** | | | | | |
| Weekday | 404.9 | 144.4 | 260.5 | 30.35 | 0.000 |
| Weekend | 370.2 | 185.3 | 184.9 | 20.31 | 0.000 |
| **Total work** | | | | | |
| Weekday | 556.1 | 572.3 | -16.2 | -1.60 | 0.109 |
| Weekend | 432.9 | 373.1 | 59.9 | 5.14 | 0.000 |
| **Leisure** | | | | | |
| Weekday | 793.5 | 791.9 | 1.6 | 0.18 | 0.857 |
| Weekend | 911.1 | 967.7 | -56.6 | -5.04 | 0.000 |
| **Leisure with children** | | | | | |
| Weekday | 192.5 | 153.2 | 39.4 | 5.48 | 0.000 |
| Weekend | 293.1 | 281.3 | 11.8 | 1.25 | 0.211 |
| **Leisure without children** | | | | | |
| Weekday | 601.0 | 638.7 | -37.7 | -5.16 | 0.000 |
| Weekend | 618.0 | 686.4 | -68.4 | -8.19 | 0.000 |
| **Childcare** | | | | | |
| Weekday | 208.8 | 99.1 | 109.7 | 17.07 | 0.000 |
| Weekend | 194.5 | 120.4 | 74.2 | 9.94 | 0.000 |
| **Housework** | | | | | |
| Weekday | 195.5 | 44.4 | 151.1 | 27.58 | 0.000 |
| Weekend | 175.3 | 64.0 | 111.2 | 19.82 | 0.000 |
| **Care for other people** | | | | | |
| Weekday | 0.5 | 0.8 | -0.4 | -0.81 | 0.417 |
| Weekend | 0.4 | 0.9 | -0.5 | -1.04 | 0.297 |
| **Total time with children** | | | | | |
| Weekday | 531.5 | 284.9 | 246.6 | 21.69 | 0.000 |
| Weekend | 635.4 | 485.9 | 149.5 | 12.35 | 0.000 |
| **Engaged time with children** | | | | | |
| Weekday | 416.7 | 240.1 | 176.7 | 17.63 | 0.000 |
| Weekend | 529.1 | 420.0 | 109.1 | 9.59 | 0.000 |
| **Quality time with children** | | | | | |
| Weekday | 303.0 | 163.7 | 139.2 | 18.35 | 0.000 |
| Weekend | 339.3 | 244.9 | 94.4 | 10.64 | 0.000 |

*Notes:* The table reports average daily time devoted to each activity, measured in minutes. The difference is computed as women minus men. t-statistics and p-values come from tests of equality of means across the two groups.

Figure 1: Leisure and the presence of children: share of leisure spent with children by gender

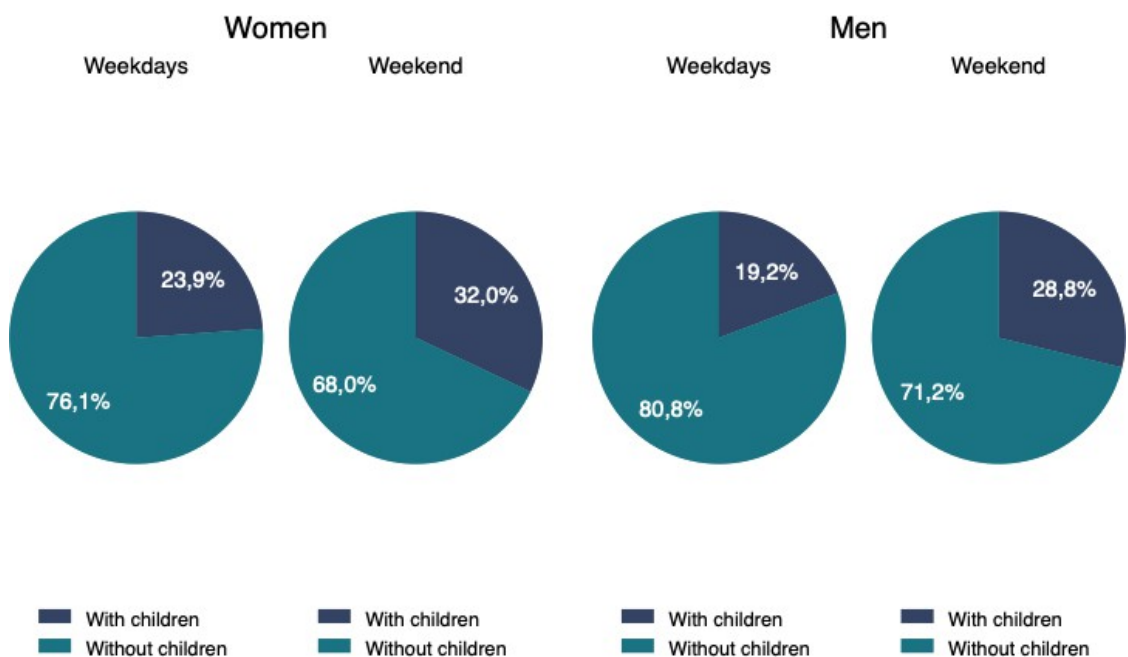

Figure 2: Leisure and the presence of children: share of leisure spent with children by gender (ER)

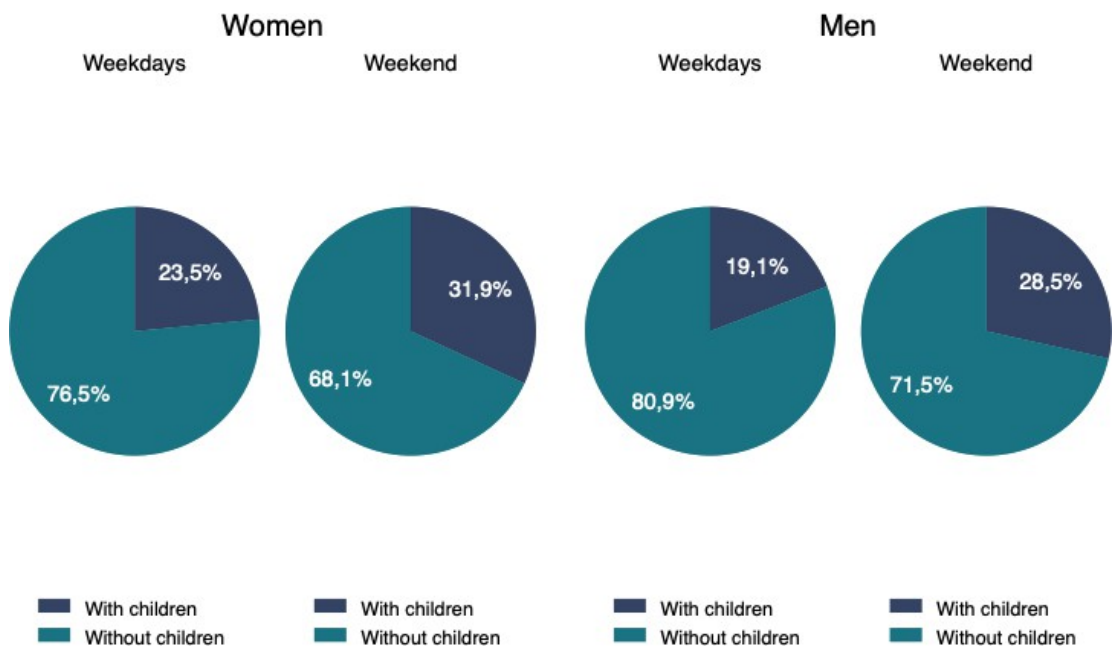

Figure 3: Leisure and the presence of children: share of leisure spent with children by gender (CA)

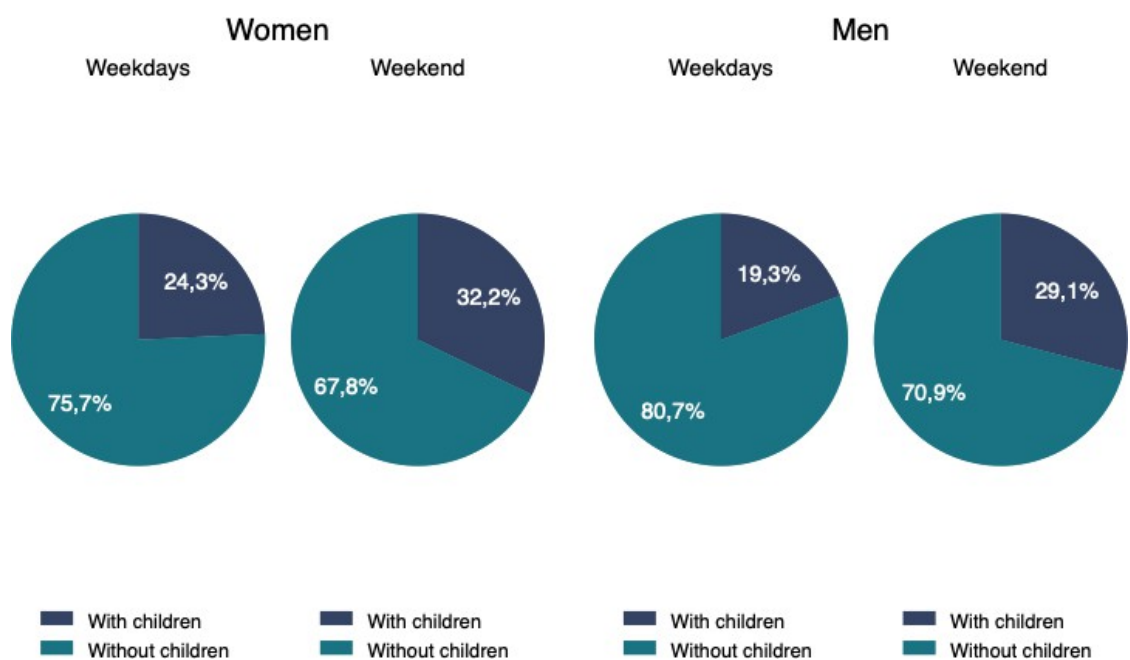

## 4.2. Gender differences in gender attitudes

The TIMES data also allow a direct comparison of men's and women's attitudes. Table 4 shows that men are, on average, more traditional than women on the dimensions that concern family roles and the male role within the couple. The sharpest differences concern statements about traditional family roles and the idea that having children is a duty. The same pattern appears in both regions, though it is generally stronger in Campania (Tables 5 and 6).

This evidence matters for two reasons. First, it shows that the time-use asymmetry documented above is embedded in a normative environment that remains gendered. Second, it suggests that the main margin of inequality is not simply women doing "too much", but men continuing to do relatively little in the domestic sphere. The descriptive comparison of attitudes is therefore consistent with the idea that domestic specialization is sustained not only by constraints but also by socially embedded expectations.

Table 4: Gender differences in attitudes toward gender norms

| Indicator | Women | Men | Diff. | t-statistic | p-value |
|---|---|---|---|---|---|
| **Traditional norms index** | 42,3 | 44,4 | -2,1 | -2,77 | 0,006 |
| Traditional family roles | 34,4 | 40,5 | -6,1 | -5,63 | 0,000 |
| Child suffers if mother works (0–6) | 57,1 | 57,8 | -0,7 | -0,66 | 0,511 |
| Child suffers if mother works (7–11) | 47,7 | 48,6 | -0,9 | -0,83 | 0,409 |
| Duty to have children | 29,1 | 34,1 | -4,9 | -4,59 | 0,000 |
| Woman earns more than man | 40,1 | 37,8 | 2,2 | 2,11 | 0,035 |
| Man takes care of the home | 38,6 | 41,2 | -2,5 | -2,43 | 0,015 |
| Woman reduces aspirations | 48,9 | 50,7 | -1,8 | -1,72 | 0,086 |

*Notes:* The table reports mean values for indicators of gender attitudes. Individual questions are measured on a 0-100 agreement scale, where higher values indicate more conservative views. Composite indices are computed as averages of the relevant items. The difference is computed as women minus men. t-statistics and p-values come from tests of equality of means across the two groups.

Table 5: Gender differences in attitudes toward gender norms (Emilia-Romagna)

| Indicator | Women | Men | Diff. | t-statistic | p-value |
|---|---|---|---|---|---|
| **Traditional norms index** | 46,0 | 47,4 | -1,4 | -1,30 | 0,194 |
| Traditional family roles | 37,8 | 41,3 | -3,5 | -2,26 | 0,024 |
| Child suffers if mother works (0–6) | 59,4 | 60,6 | -1,2 | -0,80 | 0,424 |
| Child suffers if mother works (7–11) | 51,4 | 52,5 | -1,0 | -0,66 | 0,513 |
| Duty to have children | 31,2 | 34,6 | -3,3 | -2,21 | 0,027 |
| Woman earns more than man | 44,0 | 41,7 | 2,3 | 1,58 | 0,115 |
| Man takes care of the home | 41,3 | 43,6 | -2,3 | -1,57 | 0,117 |
| Woman reduces aspirations | 56,8 | 57,6 | -0,8 | -0,55 | 0,584 |

*Notes:* The table reports mean values for indicators of gender attitudes. Individual questions are measured on a 0-100 agreement scale, where higher values indicate more conservative views. Composite indices are computed as averages of the relevant items. The difference is computed as women minus men. t-statistics and p-values come from tests of equality of means across the two groups.

Table 6: Gender differences in attitudes toward gender norms (Campania)

| Indicator | Women | Men | Diff. | t-statistic | p-value |
|---|---|---|---|---|---|
| **Traditional norms index** | 38,7 | 41,5 | -2,8 | -2,65 | 0,008 |
| Traditional family roles | 31,1 | 39,7 | -8,7 | -5,71 | 0,000 |
| Child suffers if mother works (0–6) | 55,0 | 55,1 | -0,2 | -0,13 | 0,899 |
| Child suffers if mother works (7–11) | 44,1 | 44,9 | -0,7 | -0,52 | 0,606 |
| Duty to have children | 27,1 | 33,6 | -6,4 | -4,26 | 0,000 |
| Woman earns more than man | 36,2 | 34,1 | 2,1 | 1,43 | 0,152 |
| Man takes care of the home | 36,0 | 38,8 | -2,8 | -1,87 | 0,062 |
| Woman reduces aspirations | 41,2 | 44,1 | -2,8 | -1,93 | 0,054 |

*Notes:* The table reports mean values for indicators of gender attitudes. Individual questions are measured on a 0-100 agreement scale, where higher values indicate more conservative views. Composite indices are computed as averages of the relevant items. The difference is computed as women minus men. t-statistics and p-values come from tests of equality of means across the two groups.

**4.3. Norms and time use: what people think and what they do**

We now turn to the relationship between gender attitudes and time allocation. The question is not whether norms can be isolated as an exogenous cause — they cannot, with these data — but whether more traditional attitudes are systematically associated with more traditional daily arrangements. The descriptive evidence suggests that they are.

Across the battery of outcomes, the most consistent pattern is that more traditional attitudes are associated with lower male involvement in unpaid work, childcare, housework, and total time with children, together with more male leisure, especially leisure without children. By contrast, the association with paid work is more heterogeneous across weekdays and weekends. The clearest descriptive signal therefore concerns the domestic sphere and the distribution of discretionary time, not a simple increase in male breadwinning.

The figures also indicate that women's and men's norm indices often point in the same direction. Although the magnitude of the associations varies, the general picture is that more traditional orientations among either partner are associated with more specialized within-couple arrangements. This is consistent with a view of norms as a shared couple-level environment rather than as purely individual opinions. It is also compatible, though not conclusive, with assortative matching on attitudes.

Importantly, the descriptive results for the subsample of couples in which both partners work full-time lead to the same substantive conclusion. Greater female labor-market attachment reduces some gaps but does not eliminate them. The unpaid-work gap remains large, especially in housework, and women continue to have less discretionary leisure. In other words, the results point less to a failure of women to enter paid work than to incomplete male adjustment in unpaid work once women do so.

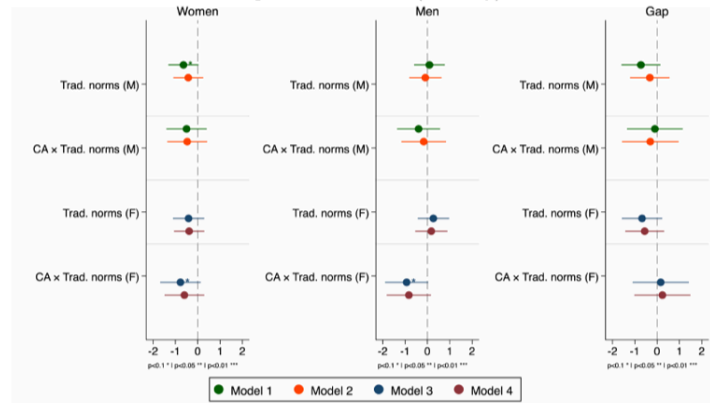

Figure 4: Paid work (weekday)

*Notes:* Each figure reports estimated coefficients and 95% confidence intervals for the main regressors and their interactions with the Campania indicator. The first panel reports estimates for women, the second for men, and the third for the within-couple gender gap, defined as female minus male. For each activity, four models are shown: (1) includes women's traditional gender-norm index; (2) adds socio-economic controls; (3) includes men's traditional gender-norm index; (4) adds socio-economic controls. Estimates are shown separately for women and men. These regressions are descriptive and should not be interpreted causally.

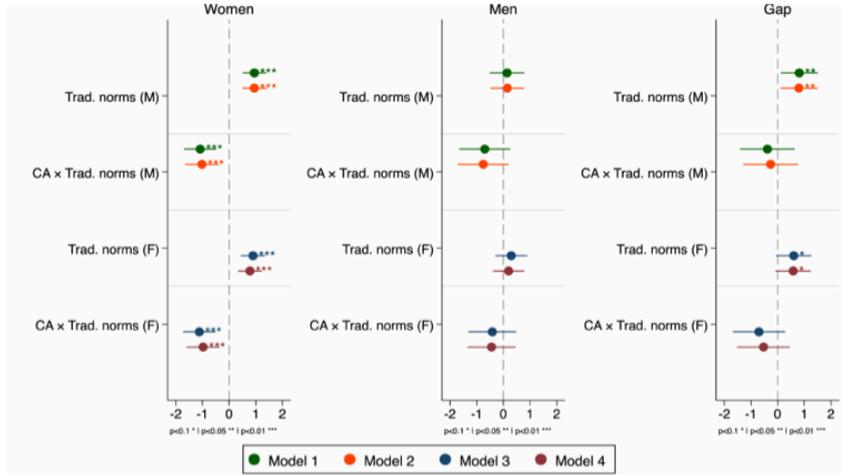

Figure 5: Paid work (weekend)

*Notes:* Each figure reports estimated coefficients and 95% confidence intervals for the main regressors and their interactions with the Campania indicator. The first panel reports estimates for women, the second for men, and the third for the within-couple gender gap, defined as female minus male. For each activity, four models are shown: (1) includes women's traditional gender-norm index; (2) adds socio-economic controls; (3) includes men's traditional gender-norm index; (4) adds socio-economic controls. Estimates are shown separately for women and men. These regressions are descriptive and should not be interpreted causally.

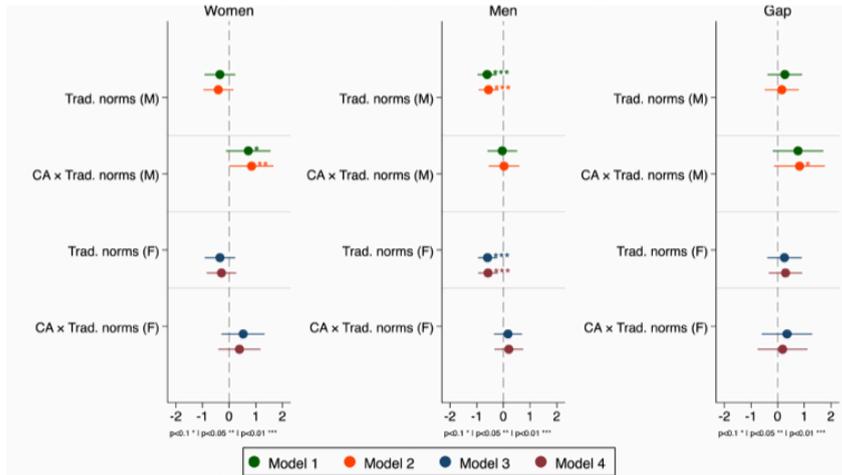

Figure 6: Unpaid work (weekday)

*Notes:* Each figure reports estimated coefficients and 95% confidence intervals for the main regressors and their interactions with the Campania indicator. The first panel reports estimates for women, the second for men, and the third for the within-couple gender gap, defined as female minus male. For each activity, four models are shown: (1) includes women's traditional gender-norm index; (2) adds socio-economic controls; (3) includes men's traditional gender-norm index; (4) adds socio-economic controls. Estimates are shown separately for women and men. These regressions are descriptive and should not be interpreted causally.

Figure 7: Unpaid work (weekend)

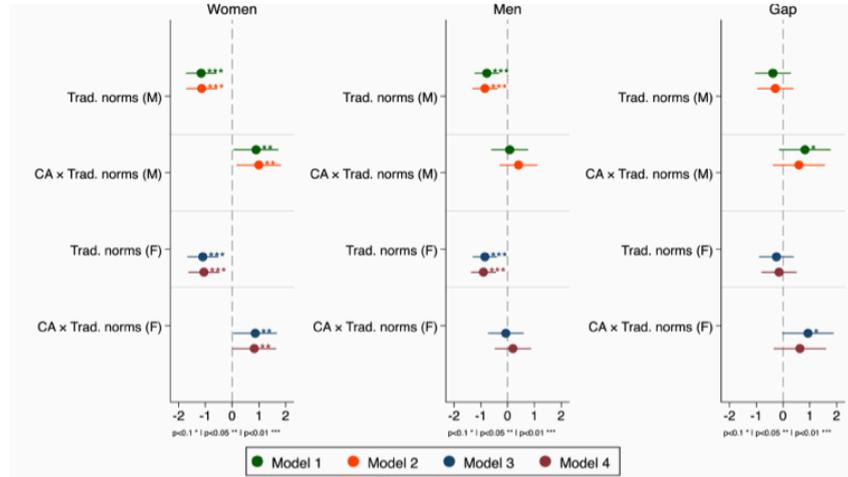

*Notes:* Each figure reports estimated coefficients and 95% confidence intervals for the main regressors and their interactions with the Campania indicator. The first panel reports estimates for women, the second for men, and the third for the within-couple gender gap, defined as female minus male. For each activity, four models are shown: (1) includes women's traditional gender-norm index; (2) adds socio-economic controls; (3) includes men's traditional gender-norm index; (4) adds socio-economic controls. Estimates are shown separately for women and men. These regressions are descriptive and should not be interpreted causally.

Figure 8: Leisure (weekday)

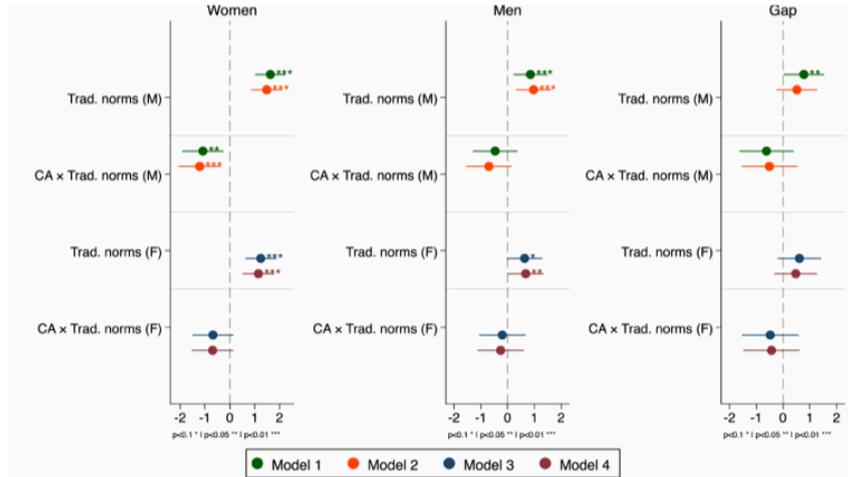

*Notes:* Each figure reports estimated coefficients and 95% confidence intervals for the main regressors and their interactions with the Campania indicator. The first panel reports estimates for women, the second for men, and the third for the within-couple gender gap, defined as female minus male. For each activity, four models are shown: (1) includes women's traditional gender-norm index; (2) adds socio-economic controls; (3) includes men's traditional gender-norm index; (4) adds socio-economic controls. Estimates are shown separately for women and men. These regressions are descriptive and should not be interpreted causally.

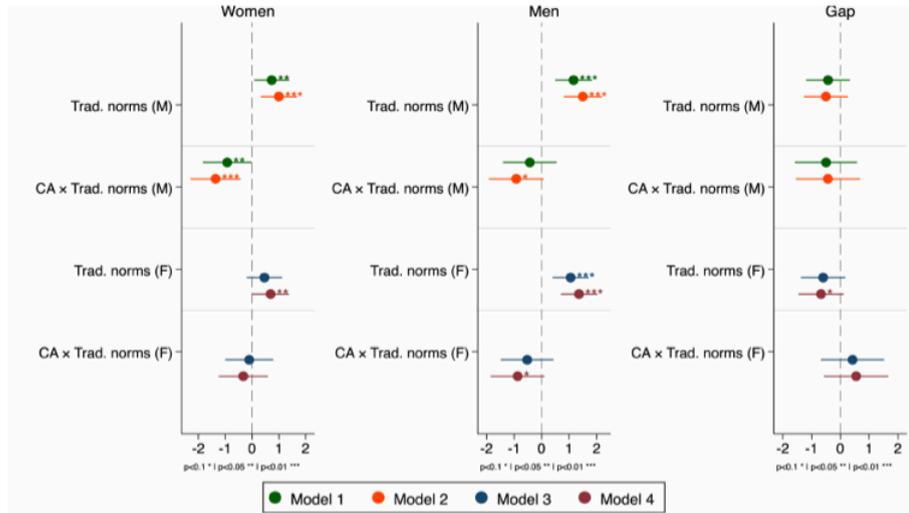

Figure 9: Leisure (weekend)

*Notes:* Each figure reports estimated coefficients and 95% confidence intervals for the main regressors and their interactions with the Campania indicator. The first panel reports estimates for women, the second for men, and the third for the within-couple gender gap, defined as female minus male. For each activity, four models are shown: (1) includes women's traditional gender-norm index; (2) adds socio-economic controls; (3) includes men's traditional gender-norm index; (4) adds socio-economic controls. Estimates are shown separately for women and men. These regressions are descriptive and should not be interpreted causally.

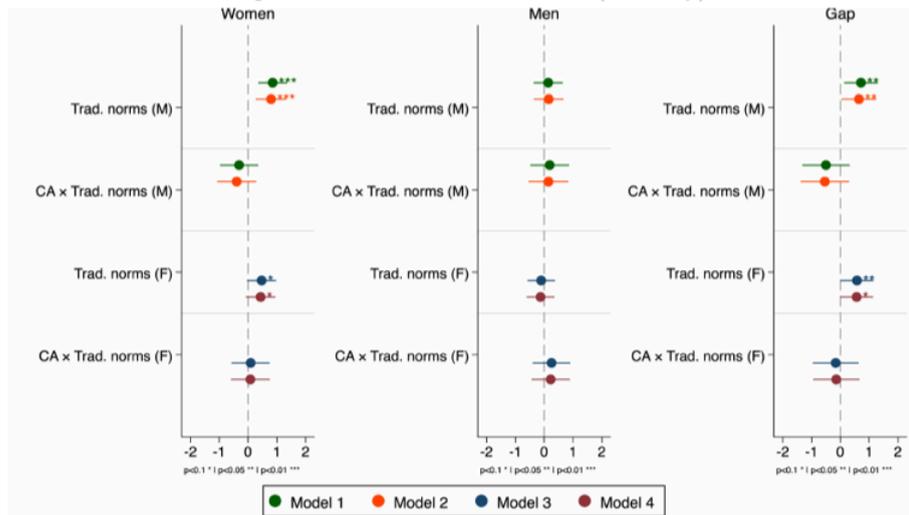

Figure 10: Leisure with children (weekday)

*Notes:* Each figure reports estimated coefficients and 95% confidence intervals for the main regressors and their interactions with the Campania indicator. The first panel reports estimates for women, the second for men, and the third for the within-couple gender gap, defined as female minus male. For each activity, four models are shown: (1) includes women's traditional gender-norm index; (2) adds socio-economic controls; (3) includes men's traditional gender-norm index; (4) adds socio-economic controls. Estimates are shown separately for women and men. These regressions are descriptive and should not be interpreted causally.

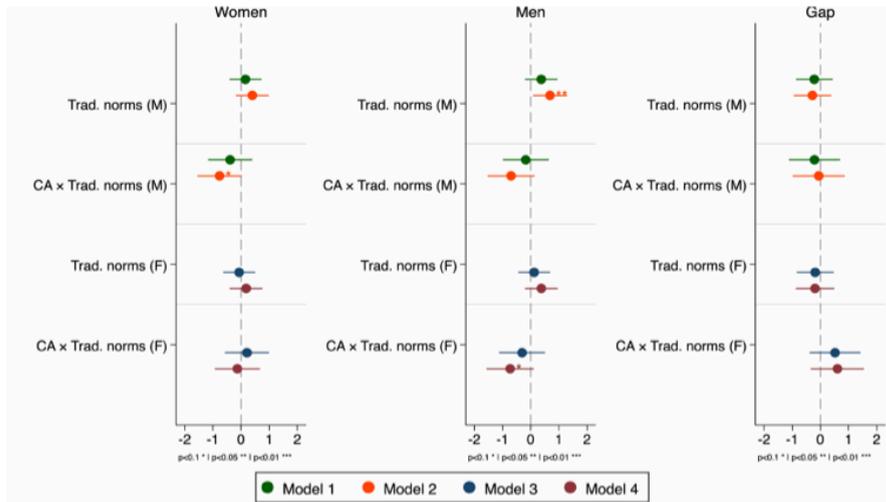

Figure 11: Leisure with children (weekend)

*Notes:* Each figure reports estimated coefficients and 95% confidence intervals for the main regressors and their interactions with the Campania indicator. The first panel reports estimates for women, the second for men, and the third for the within-couple gender gap, defined as female minus male. For each activity, four models are shown: (1) includes women's traditional gender-norm index; (2) adds socio-economic controls; (3) includes men's traditional gender-norm index; (4) adds socio-economic controls. Estimates are shown separately for women and men. These regressions are descriptive and should not be interpreted causally.

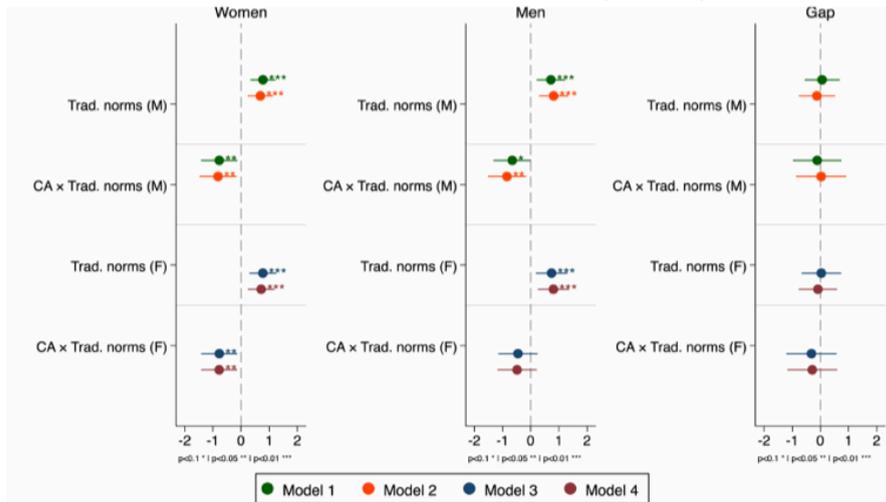

Figure 12: Leisure without children (weekday)

*Notes:* Each figure reports estimated coefficients and 95% confidence intervals for the main regressors and their interactions with the Campania indicator. The first panel reports estimates for women, the second for men, and the third for the within-couple gender gap, defined as female minus male. For each activity, four models are shown: (1) includes women's traditional gender-norm index; (2) adds socio-economic controls; (3) includes men's traditional gender-norm index; (4) adds socio-economic controls. Estimates are shown separately for women and men. These regressions are descriptive and should not be interpreted causally.

Figure 13: Leisure without children (weekend)

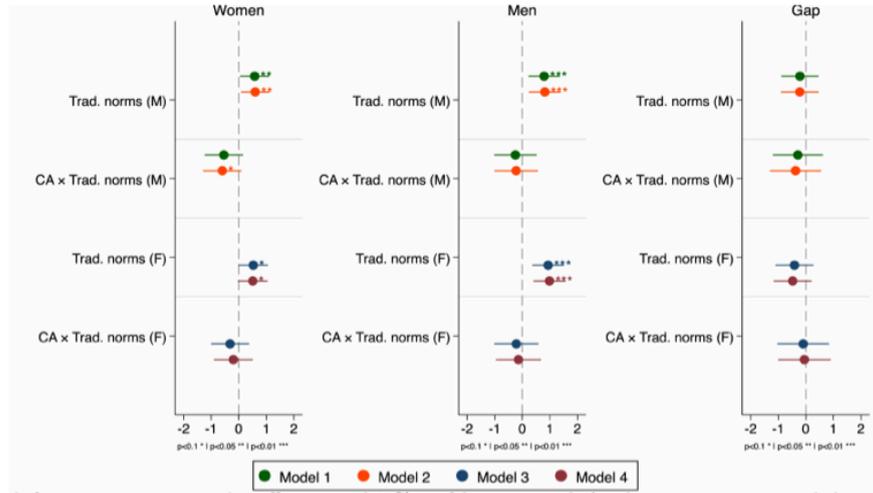

*Notes:* Each figure reports estimated coefficients and 95% confidence intervals for the main regressors and their interactions with the Campania indicator. The first panel reports estimates for women, the second for men, and the third for the within-couple gender gap, defined as female minus male. For each activity, four models are shown: (1) includes women's traditional gender-norm index; (2) adds socio-economic controls; (3) includes men's traditional gender-norm index; (4) adds socio-economic controls. Estimates are shown separately for women and men. These regressions are descriptive and should not be interpreted causally.

Figure 14: Childcare (weekday)

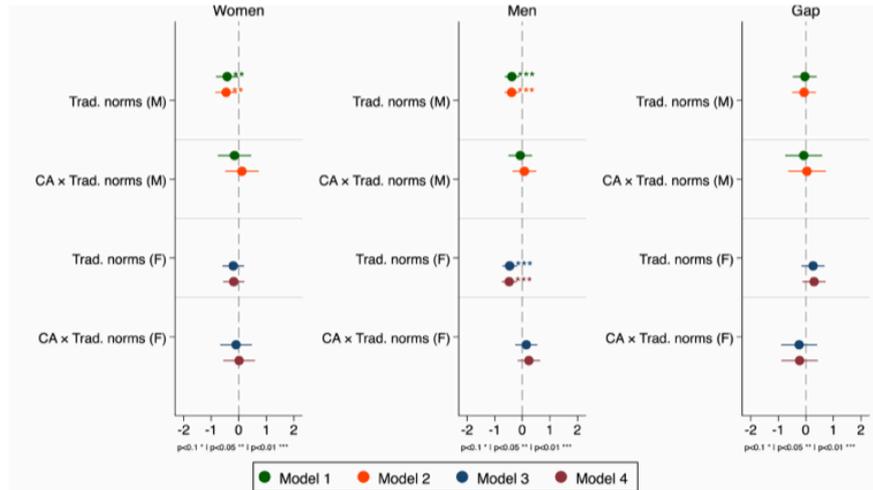

*Notes:* Each figure reports estimated coefficients and 95% confidence intervals for the main regressors and their interactions with the Campania indicator. The first panel reports estimates for women, the second for men, and the third for the within-couple gender gap, defined as female minus male. For each activity, four models are shown: (1) includes women's traditional gender-norm index; (2) adds socio-economic controls; (3) includes men's traditional gender-norm index; (4) adds socio-economic controls. Estimates are shown separately for women and men. These regressions are descriptive and should not be interpreted causally.

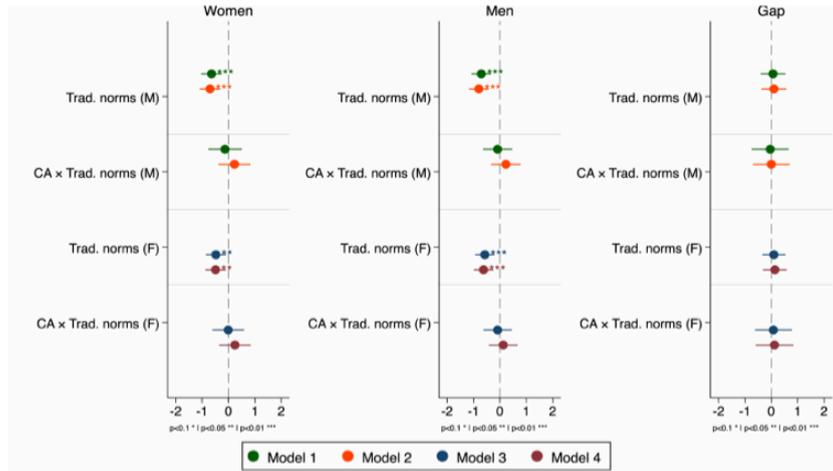

Figure 15: Childcare (weekend)

*Notes:* Each figure reports estimated coefficients and 95% confidence intervals for the main regressors and their interactions with the Campania indicator. The first panel reports estimates for women, the second for men, and the third for the within-couple gender gap, defined as female minus male. For each activity, four models are shown: (1) includes women's traditional gender-norm index; (2) adds socio-economic controls; (3) includes men's traditional gender-norm index; (4) adds socio-economic controls. Estimates are shown separately for women and men. These regressions are descriptive and should not be interpreted causally.

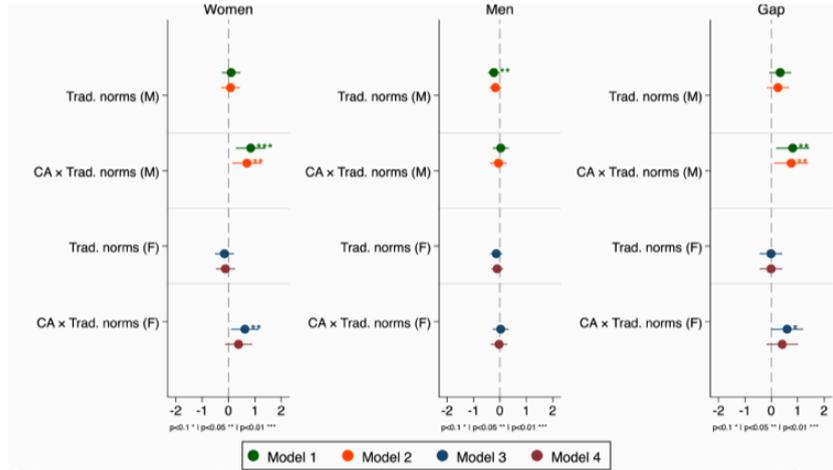

Figure 16: Housework (weekday)

*Notes:* Each figure reports estimated coefficients and 95% confidence intervals for the main regressors and their interactions with the Campania indicator. The first panel reports estimates for women, the second for men, and the third for the within-couple gender gap, defined as female minus male. For each activity, four models are shown: (1) includes women's traditional gender-norm index; (2) adds socio-economic controls; (3) includes men's traditional gender-norm index; (4) adds socio-economic controls. Estimates are shown separately for women and men. These regressions are descriptive and should not be interpreted causally.

Figure 17: Housework (weekend)

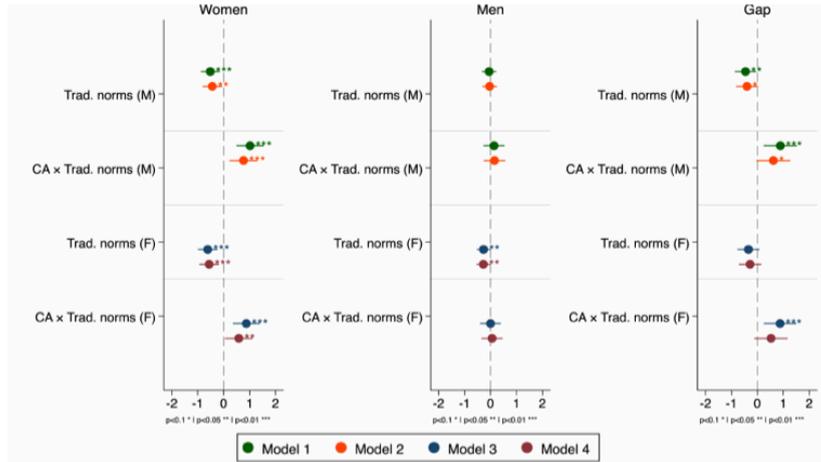

*Notes:* Each figure reports estimated coefficients and 95% confidence intervals for the main regressors and their interactions with the Campania indicator. The first panel reports estimates for women, the second for men, and the third for the within-couple gender gap, defined as female minus male. For each activity, four models are shown: (1) includes women's traditional gender-norm index; (2) adds socio-economic controls; (3) includes men's traditional gender-norm index; (4) adds socio-economic controls. Estimates are shown separately for women and men. These regressions are descriptive and should not be interpreted causally.

Figure 18: Total work (weekday)

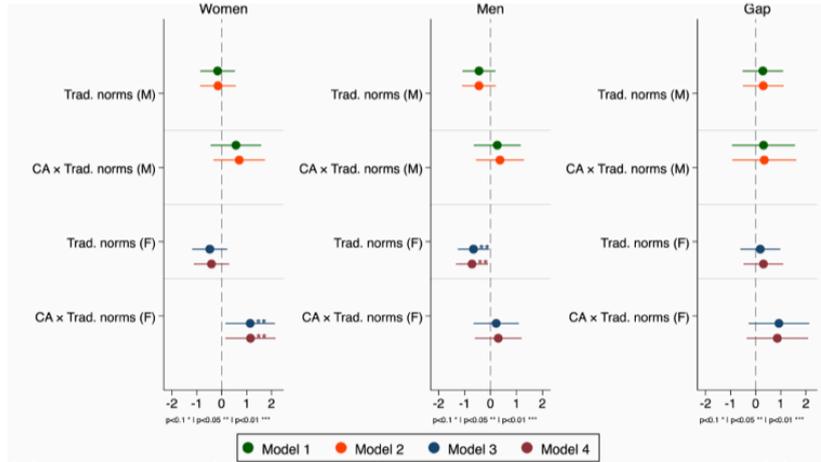

*Notes:* Each figure reports estimated coefficients and 95% confidence intervals for the main regressors and their interactions with the Campania indicator. The first panel reports estimates for women, the second for men, and the third for the within-couple gender gap, defined as female minus male. For each activity, four models are shown: (1) includes women's traditional gender-norm index; (2) adds socio-economic controls; (3) includes men's traditional gender-norm index; (4) adds socio-economic controls. Estimates are shown separately for women and men. These regressions are descriptive and should not be interpreted causally.

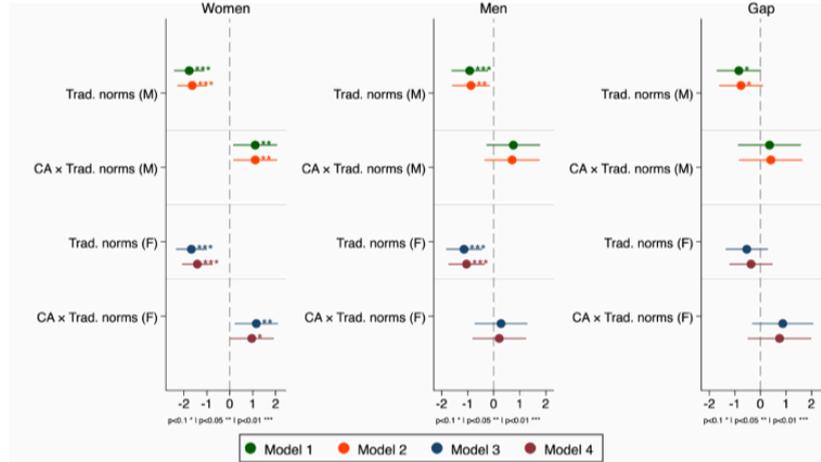

Figure 19: Total work (weekend)

*Notes:* Each figure reports estimated coefficients and 95% confidence intervals for the main regressors and their interactions with the Campania indicator. The first panel reports estimates for women, the second for men, and the third for the within-couple gender gap, defined as female minus male. For each activity, four models are shown: (1) includes women's traditional gender-norm index; (2) adds socio-economic controls; (3) includes men's traditional gender-norm index; (4) adds socio-economic controls. Estimates are shown separately for women and men. These regressions are descriptive and should not be interpreted causally.

## 5. Discussion and policy implications

The evidence helps resolve the paper's opening puzzle. The persistence of gender inequality among young Italian families is not mainly a story of women failing to enter paid work. It is a story of incomplete convergence inside the household once women do so. On weekdays, women and men often perform similar amounts of total work, but the composition of that work remains sharply asymmetric. Women continue to absorb most unpaid work and spend more time with children, while men retain greater access to leisure without children. On weekends, these asymmetries remain visible and often widen.

This distinction between total work and its composition matters for how gender inequality is interpreted. If women and men perform similar total work but one partner bears more unpaid work and has less discretionary leisure, then equality in total hours does not imply equality in lived well-being. The results therefore suggest that within-household inequality should be measured using a broader family of indicators: not only paid and unpaid work, but also constrained versus unconstrained leisure, time with children, and responsibility for the organization of family life.

The attitudinal evidence points in the same direction. More traditional views are systematically associated with lower male domestic participation and wider male access to unconstrained leisure. The strongest descriptive relationships are not in women's paid work but in men's behavior inside the household. That pattern is consistent with a slow redefinition of masculine family roles and with the idea that norms help stabilize unequal arrangements even when women's labor-market attachment increases. The paper cannot establish whether norms cause behavior, whether behavior rationalizes norms, or whether both are jointly determined. But it does show that the two are empirically intertwined in ways that matter for everyday life.

The regional comparison adds a further layer. Campania exhibits wider time gaps and more polarized attitudes than Emilia-Romagna, suggesting that labor-market conditions, care infrastructures, and the normative environment are mutually reinforcing. In more constrained contexts, the family absorbs more unpaid work and gender specialization becomes harder to break. This is consistent with a broader literature that treats work-family reconciliation as jointly shaped by institutions and social norms rather than by household preferences alone.

Two qualifications are important. First, the paper studies a specific population: co-resident couples with at least one child younger than 11 in Emilia-Romagna or Campania. The results should therefore be interpreted as evidence on the organization of time within young families in these settings, not as claims about all Italian households. Second, the results linking attitudes to time allocation are informative because they locate the strongest associations in men's unpaid work and leisure without children, but they remain associational.

These qualifications do not reduce the paper's policy relevance. If the objective is a more equal distribution of well-being within families, increasing female employment is necessary but not sufficient. Without a parallel increase in male involvement in unpaid work, stronger female labor-market attachment may intensify women's double burden rather than reduce inequality. Policies that expand childcare availability, full-time schooling, and after-school services remain important because they relax material constraints. But policies that explicitly target men's involvement—such as non-transferable paternity leave, workplace arrangements that legitimize fathers' care responsibilities, and public communication that challenges traditional role expectations—are equally important. From a social-indicators perspective, progress should be monitored not only through labor-market participation, but also through indicators of domestic redistribution and discretionary time.

## 6. Conclusion

This paper has examined how paid work, unpaid work, childcare, leisure, and gender attitudes are organized within young Italian families. It began from a puzzle: why do large gender inequalities persist in everyday life even as women strengthen their attachment to paid work? Using matched partner diaries and direct measures of gender attitudes, the paper shows that the answer lies in the incomplete redistribution of unpaid work and discretionary time within couples.

Three conclusions follow. First, within-couple gender gaps in unpaid work remain very large, and women also spend more time with children. Second, those gaps persist even among dual full-time couples, implying that stronger female labor-market attachment is not sufficient on its own to equalize daily life. Third, more traditional gender attitudes are consistently associated with lower male domestic involvement and more unequal access to leisure without children.

The paper's main contribution is to fill a measurement gap. By combining matched partner diaries, socio-economic information, and direct indicators of gender norms, the TIMES Observatory makes it possible to observe dimensions of family inequality that standard surveys usually miss. In that sense, the paper proposes and applies a set of couple-level indicators that help reveal how gender inequality is reproduced through the daily organization of work, care, and autonomy. These are not causal estimates, but they provide a clearer empirical basis for understanding inequality in family well-being and for monitoring whether change is occurring where it matters most: inside households.

The broader takeaway is that gender inequality within couples is not only a matter of who works more in total, but of who has access to time that is genuinely discretionary. Women may have less time that is continuous, less interruptible, and more available for self-directed use, even when aggregate differences in leisure appear modest. Seen in this way, time inequality is also inequality in autonomy. 

# A. Additional descriptive tables

Table 7: Socio-economic descriptive statistics

|  | N | Mean | SD | Min | Max |
|---|---|---|---|---|---|
| **Women** | | | | | |
| Age | 1.928 | 37,42 | 6,91 | 19,00 | 61,00 |
| Number of children | 1.928 | 1,27 | 0,49 | 1,00 | 5,00 |
| Age of oldest child | 1.928 | 5,10 | 3,15 | 0,00 | 11,00 |
| Employed | 1.928 | 0,52 | 0,50 | 0,00 | 1,00 |
| Employed full-time | 999 | 0,60 | 0,49 | 0,00 | 1,00 |
| College degree or more | 1.864 | 0,40 | 0,49 | 0,00 | 1,00 |
| Resident in Emilia-Romagna | 1.928 | 0,49 | 0,50 | 0,00 | 1,00 |
| **Men** | | | | | |
| Age | 1.928 | 39,92 | 7,74 | 18,00 | 81,00 |
| Number of children | 1.928 | 1,27 | 0,49 | 1,00 | 5,00 |
| Age of oldest child | 1.928 | 5,10 | 3,15 | 0,00 | 11,00 |
| Employed | 1.928 | 0,95 | 0,21 | 0,00 | 1,00 |
| Employed full-time | 1.838 | 0,80 | 0,40 | 0,00 | 1,00 |
| College degree or more | 1.855 | 0,38 | 0,48 | 0,00 | 1,00 |
| Resident in Emilia-Romagna | 1.928 | 0,49 | 0,50 | 0,00 | 1,00 |

*Notes:* The table reports socio-economic descriptive statistics for women and men. The variable *Employed full-time* is defined only among employed respondents. The number of observations for *College degree or more* is lower because some respondents did not provide information on education.

Table 8: Socio-economic descriptive statistics (Emilia-Romagna)

|  | N | Mean | SD | Min | Max |
|---|---|---|---|---|---|
| **Women** | | | | | |
| Age | 951 | 37,41 | 7,58 | 19,00 | 60,00 |
| Number of children | 951 | 1,24 | 0,47 | 1,00 | 5,00 |
| Age of oldest child | 951 | 5,19 | 3,18 | 0,00 | 10,00 |
| Employed | 951 | 0,69 | 0,46 | 0,00 | 1,00 |
| Employed full-time | 653 | 0,68 | 0,47 | 0,00 | 1,00 |
| College degree or more | 928 | 0,38 | 0,49 | 0,00 | 1,00 |
| **Men** | | | | | |
| Age | 951 | 39,83 | 8,47 | 18,00 | 81,00 |
| Number of children | 951 | 1,24 | 0,47 | 1,00 | 5,00 |
| Age of oldest child | 951 | 5,19 | 3,18 | 0,00 | 10,00 |
| Employed | 951 | 0,98 | 0,15 | 0,00 | 1,00 |
| Employed full-time | 928 | 0,85 | 0,35 | 0,00 | 1,00 |
| College degree or more | 927 | 0,43 | 0,50 | 0,00 | 1,00 |

*Notes:* The table reports socio-economic descriptive statistics for women and men. The variable *Employed full-time* is defined only among employed respondents. The number of observations for *College degree or more* is lower because some respondents did not provide information on education.

|  | N | Mean | SD | Min | Max |
|---|---|---|---|---|---|
| **Women** | | | | | |
| Age | 977 | 37,43 | 6,19 | 20,00 | 61,00 |
| Number of children | 977 | 1,29 | 0,51 | 1,00 | 4,00 |
| Age of oldest child | 977 | 5,01 | 3,11 | 0,00 | 11,00 |
| Employed | 977 | 0,35 | 0,48 | 0,00 | 1,00 |
| Employed full-time | 346 | 0,45 | 0,50 | 0,00 | 1,00 |
| College degree or more | 936 | 0,41 | 0,49 | 0,00 | 1,00 |
| **Men** | | | | | |
| Age | 977 | 40,01 | 6,96 | 20,00 | 64,00 |
| Number of children | 977 | 1,29 | 0,51 | 1,00 | 4,00 |
| Age of oldest child | 977 | 5,01 | 3,11 | 0,00 | 11,00 |
| Employed | 977 | 0,93 | 0,25 | 0,00 | 1,00 |
| Employed full-time | 910 | 0,74 | 0,44 | 0,00 | 1,00 |
| College degree or more | 928 | 0,32 | 0,47 | 0,00 | 1,00 |

*Notes:* The table reports socio-economic descriptive statistics for women and men. The variable *Employed full-time* is defined only among employed respondents. The number of observations for *College degree or more* is lower because some respondents did not provide information on education.

Table 10: Descriptive statistics for time use

|  | N | Mean | SD | Min | Max |
|---|---|---|---|---|---|
| **Paid work** | | | | | |
| Women (Weekday) | 1,928 | 207.91 | 246.44 | 0.00 | 1110.00 |
| Men (Weekday) | 1,928 | 427.08 | 250.48 | 0.00 | 1110.00 |
| Gap (Weekday) | 1,928 | -219.16 | 325.34 | -1110.00 | 870.00 |
| Women (Weekend) | 1,928 | 64.39 | 167.61 | 0.00 | 1110.00 |
| Men (Weekend) | 1,928 | 151.38 | 248.40 | 0.00 | 1060.00 |
| Gap (Weekend) | 1,928 | -86.99 | 267.52 | -1060.00 | 1110.00 |
| **Unpaid work** | | | | | |
| Women (Weekday) | 1,928 | 339.38 | 223.91 | 0.00 | 1290.00 |
| Men (Weekday) | 1,928 | 128.84 | 139.56 | 0.00 | 1050.00 |
| Gap (Weekday) | 1,928 | 210.53 | 253.17 | -730.00 | 1220.00 |
| Women (Weekend) | 1,928 | 325.07 | 216.37 | 0.00 | 1390.00 |
| Men (Weekend) | 1,928 | 187.43 | 173.93 | 0.00 | 1020.00 |
| Gap (Weekend) | 1,928 | 137.65 | 251.36 | -900.00 | 1050.00 |
| **Total work** | | | | | |
| Women (Weekday) | 1,928 | 547.29 | 231.58 | 0.00 | 1290.00 |
| Men (Weekday) | 1,928 | 555.92 | 238.19 | 0.00 | 1200.00 |
| Gap (Weekday) | 1,928 | -8.63 | 283.07 | -950.00 | 1110.00 |
| Women (Weekend) | 1,928 | 389.46 | 242.68 | 0.00 | 1390.00 |
| Men (Weekend) | 1,928 | 338.81 | 262.39 | 0.00 | 1220.00 |
| Gap (Weekend) | 1,928 | 50.65 | 285.10 | -1070.00 | 1230.00 |
| **Leisure** | | | | | |
| Women (Weekday) | 1,928 | 815.47 | 221.90 | 0.00 | 1440.00 |
| Men (Weekday) | 1,928 | 815.70 | 220.98 | 150.00 | 1440.00 |
| Gap (Weekday) | 1,928 | -0.23 | 262.09 | -1020.00 | 990.00 |
| Women (Weekend) | 1,928 | 952.36 | 240.80 | 50.00 | 1440.00 |
| Men (Weekend) | 1,928 | 1002.60 | 259.03 | 10.00 | 1440.00 |
| Gap (Weekend) | 1,928 | -50.24 | 274.71 | -1230.00 | 1310.00 |
| **Childcare** | | | | | |
| Women (Weekday) | 1,928 | 174.80 | 155.52 | 0.00 | 1170.00 |
| Men (Weekday) | 1,928 | 88.06 | 103.59 | 0.00 | 1020.00 |
| Gap (Weekday) | 1,928 | 86.74 | 169.84 | -620.00 | 1010.00 |
| Women (Weekend) | 1,928 | 164.76 | 171.05 | 0.00 | 1180.00 |
| Men (Weekend) | 1,928 | 117.87 | 138.87 | 0.00 | 1020.00 |
| Gap (Weekend) | 1,928 | 46.89 | 191.71 | -850.00 | 1080.00 |
| **Housework** | | | | | |
| Women (Weekday) | 1,928 | 163.99 | 144.59 | 0.00 | 1280.00 |
| Men (Weekday) | 1,928 | 40.31 | 81.20 | 0.00 | 780.00 |
| Gap (Weekday) | 1,928 | 123.68 | 165.23 | -610.00 | 1280.00 |
| Women (Weekend) | 1,928 | 160.04 | 134.86 | 0.00 | 900.00 |
| Men (Weekend) | 1,928 | 68.87 | 107.68 | 0.00 | 1020.00 |
| Gap (Weekend) | 1,928 | 91.16 | 162.79 | -830.00 | 900.00 |
| **Total time with children** | | | | | |
| Women (Weekday) | 1,928 | 479.14 | 269.26 | 0.00 | 1440.00 |
| Men (Weekday) | 1,928 | 620.93 | 254.67 | 0.00 | 1440.00 |
| Gap (Weekday) | | | | | |
| Women (Weekend) | 1,928 | 386.97 | 232.23 | 0.00 | 1320.00 |
| Men (Weekend) | 1,928 | 526.87 | 244.23 | 0.00 | 1440.00 |
| Gap (Weekend) | | | | | |
| **Engaged time with children** | | | | | |
| Women (Weekday) | 1,928 | 324.76 | 199.59 | 0.00 | 1080.00 |
| Men (Weekday) | | | | | |
| Gap (Weekday) | 1,928 | 278.25 | 228.18 | 0.00 | 1240.00 |
| Women (Weekend) | 1,928 | 499.70 | 275.94 | 0.00 | 1380.00 |
| Men (Weekend) | 1,928 | 240.62 | 208.26 | 0.00 | 1240.00 |
| Gap (Weekend) | 1,928 | 441.50 | 266.36 | 0.00 | 1380.00 |
| **Quality time with children** | | | | | |
| Women (Weekday) | 1,928 | 247.85 | 185.70 | 0.00 | 960.00 |
| Men (Weekday) | 1,928 | 109.91 | 217.53 | -810.00 | 1200.00 |
| Gap (Weekday) | 1,928 | 76.91 | 241.62 | -780.00 | 960.00 |
| Women (Weekend) | 1,928 | 146.35 | 273.97 | -1080.00 | 1240.00 |
| Men (Weekend) | 1,928 | 85.37 | 297.33 | -1100.00 | 1230.00 |
| Gap (Weekend) | 1,928 | 200.89 | 316.30 | -1140.00 | 1320.00 |
|  | 1,928 | 121.23 | 319.79 | -960.00 | 1230.00 |

*Notes:* The table reports descriptive statistics for time-use variables, measured in minutes per day.

Table 11: Descriptive statistics for time use (Emilia-Romagna)

|  | N | Mean | SD | Min | Max |
|---|---|---|---|---|---|
| **Paid work** | | | | | |
| Women (Weekday) | 951 | 266.10 | 260.47 | 0.00 | 870.00 |
| Men (Weekday) | 951 | 426.20 | 255.39 | 0.00 | 1080.00 |
| Gap (Weekday) | 951 | -160.11 | 324.87 | -1080.00 | 870.00 |
| Women (Weekend) | 951 | 66.07 | 174.45 | 0.00 | 960.00 |
| Men (Weekend) | 951 | 113.95 | 226.07 | 0.00 | 960.00 |
| Gap (Weekend) | 951 | -47.89 | 244.60 | -900.00 | 900.00 |
| **Unpaid work** | | | | | |
| Women (Weekday) | 951 | 272.10 | 206.88 | 0.00 | 1140.00 |
| Men (Weekday) | 951 | 112.88 | 122.61 | 0.00 | 750.00 |
| Gap (Weekday) | 951 | 159.22 | 232.72 | -650.00 | 960.00 |
| Women (Weekend) | 951 | 278.72 | 200.37 | 0.00 | 1390.00 |
| Men (Weekend) | 951 | 189.66 | 169.36 | 0.00 | 900.00 |
| Gap (Weekend) | 951 | 89.05 | 232.92 | -900.00 | 840.00 |
| **Total work** | | | | | |
| Women (Weekday) | 951 | 538.20 | 241.74 | 0.00 | 1140.00 |
| Men (Weekday) | 951 | 539.09 | 249.95 | 0.00 | 1200.00 |
| Gap (Weekday) | 951 | -0.88 | 290.88 | -950.00 | 900.00 |
| Women (Weekend) | 951 | 344.78 | 235.87 | 0.00 | 1390.00 |
| Men (Weekend) | 951 | 303.62 | 246.30 | 0.00 | 1020.00 |
| Gap (Weekend) | 951 | 41.17 | 278.26 | -900.00 | 900.00 |
| **Leisure** | | | | | |
| Women (Weekday) | 951 | 838.04 | 237.79 | 210.00 | 1440.00 |
| Men (Weekday) | 951 | 840.19 | 240.59 | 210.00 | 1440.00 |
| Gap (Weekday) | 951 | -2.15 | 276.68 | -950.00 | 990.00 |
| Women (Weekend) | 951 | 994.75 | 238.79 | 50.00 | 1440.00 |
| Men (Weekend) | 951 | 1038.45 | 252.63 | 310.00 | 1440.00 |
| Gap (Weekend) | 951 | -43.70 | 261.67 | -900.00 | 830.00 |
| **Childcare** | | | | | |
| Women (Weekday) | 951 | 139.83 | 136.98 | 0.00 | 930.00 |
| Men (Weekday) | 951 | 76.70 | 89.12 | 0.00 | 600.00 |
| Gap (Weekday) | 951 | 63.13 | 150.86 | -540.00 | 840.00 |
| Women (Weekend) | 951 | 134.19 | 147.26 | 0.00 | 780.00 |
| Men (Weekend) | 951 | 115.31 | 138.00 | 0.00 | 720.00 |
| Gap (Weekend) | 951 | 18.87 | 174.64 | -690.00 | 750.00 |
| **Housework** | | | | | |
| Women (Weekday) | 951 | 131.58 | 133.71 | 0.00 | 690.00 |
| Men (Weekday) | 951 | 36.10 | 75.23 | 0.00 | 690.00 |
| Gap (Weekday) | 951 | 95.48 | 153.54 | -520.00 | 690.00 |
| Women (Weekend) | 951 | 144.38 | 130.64 | 0.00 | 810.00 |
| Men (Weekend) | 951 | 73.86 | 105.74 | 0.00 | 830.00 |
| Gap (Weekend) | 951 | 70.53 | 150.39 | -830.00 | 810.00 |
| **Total time with children** | | | | | |
| Women (Weekday) | 951 | 425.35 | 259.56 | 0.00 | 1170.00 |
| Men (Weekday) | 951 | 606.05 | 257.20 | 0.00 | 1440.00 |
| Gap (Weekday) | | | | | |
| Women (Weekend) | 951 | 356.38 | 228.03 | 0.00 | 1170.00 |
| Men (Weekend) | 951 | 524.61 | 252.58 | 0.00 | 1440.00 |
| Gap (Weekend) | | | | | |
| **Engaged time with children** | | | | | |
| Women (Weekday) | 951 | 309.80 | 199.28 | 0.00 | 840.00 |
| Men (Weekday) | | | | | |
| Gap (Weekday) | 951 | 271.38 | 223.24 | 0.00 | 1240.00 |
| Women (Weekend) | 951 | 513.87 | 268.14 | 0.00 | 1190.00 |
| Men (Weekend) | 951 | 241.19 | 206.66 | 0.00 | 1240.00 |
| Gap (Weekend) | 951 | 463.58 | 265.31 | 0.00 | 1170.00 |
| **Quality time with children** | | | | | |
| Women (Weekday) | 951 | 250.86 | 177.78 | 0.00 | 840.00 |
| Men (Weekday) | 951 | 79.78 | 198.41 | -780.00 | 870.00 |
| Gap (Weekday) | 951 | 58.94 | 230.80 | -780.00 | 800.00 |
| Women (Weekend) | 951 | 115.19 | 261.22 | -990.00 | 1130.00 |
| Men (Weekend) | 951 | 61.03 | 290.46 | -1100.00 | 950.00 |
| Gap (Weekend) | 951 | 153.97 | 289.30 | -990.00 | 1090.00 |
| | 951 | 92.18 | 303.92 | -960.00 | 1160.00 |

*Notes:* The table reports descriptive statistics for time-use variables, measured in minutes per day.

Table 12: Descriptive statistics for time use (Campania)

|  | N | Mean | SD | Min | Max |
|---|---|---|---|---|---|
| **Paid work** | | | | | |
| Women (Weekday) | 977 | 151.28 | 217.64 | 0.00 | 1110.00 |
| Men (Weekday) | 977 | 427.93 | 245.74 | 0.00 | 1110.00 |
| Gap (Weekday) | 977 | -276.65 | 315.51 | -1110.00 | 720.00 |
| Women (Weekend) | 977 | 62.75 | 160.75 | 0.00 | 1110.00 |
| Men (Weekend) | 977 | 187.81 | 263.41 | 0.00 | 1060.00 |
| Gap (Weekend) | 977 | -125.06 | 283.06 | -1060.00 | 1110.00 |
| **Unpaid work** | | | | | |
| Women (Weekday) | 977 | 404.86 | 220.52 | 0.00 | 1290.00 |
| Men (Weekday) | 977 | 144.38 | 152.75 | 0.00 | 1050.00 |
| Gap (Weekday) | 977 | 260.48 | 262.25 | -730.00 | 1220.00 |
| Women (Weekend) | 977 | 370.19 | 221.86 | 0.00 | 1320.00 |
| Men (Weekend) | 977 | 185.25 | 178.33 | 0.00 | 1020.00 |
| Gap (Weekend) | 977 | 184.94 | 259.61 | -860.00 | 1050.00 |
| **Total work** | | | | | |
| Women (Weekday) | 977 | 556.14 | 221.01 | 0.00 | 1290.00 |
| Men (Weekday) | 977 | 572.31 | 225.09 | 0.00 | 1160.00 |
| Gap (Weekday) | 977 | -16.17 | 275.20 | -750.00 | 1110.00 |
| Women (Weekend) | 977 | 432.95 | 241.42 | 0.00 | 1350.00 |
| Men (Weekend) | 977 | 373.07 | 272.96 | 0.00 | 1220.00 |
| Gap (Weekend) | 977 | 59.88 | 291.44 | -1070.00 | 1230.00 |
| **Leisure** | | | | | |
| Women (Weekday) | 977 | 793.50 | 202.97 | 0.00 | 1440.00 |
| Men (Weekday) | 977 | 791.86 | 197.29 | 150.00 | 1440.00 |
| Gap (Weekday) | 977 | 1.64 | 247.20 | -1020.00 | 750.00 |
| Women (Weekend) | 977 | 911.10 | 235.65 | 90.00 | 1440.00 |
| Men (Weekend) | 977 | 967.69 | 260.55 | 10.00 | 1440.00 |
| Gap (Weekend) | 977 | -56.60 | 286.83 | -1230.00 | 1310.00 |
| **Childcare** | | | | | |
| Women (Weekday) | 977 | 208.84 | 164.74 | 0.00 | 1170.00 |
| Men (Weekday) | 977 | 99.12 | 114.92 | 0.00 | 1020.00 |
| Gap (Weekday) | 977 | 109.72 | 183.65 | -620.00 | 1010.00 |
| Women (Weekend) | 977 | 194.52 | 186.71 | 0.00 | 1180.00 |
| Men (Weekend) | 977 | 120.36 | 139.75 | 0.00 | 1020.00 |
| Gap (Weekend) | 977 | 74.17 | 203.39 | -850.00 | 1080.00 |
| **Housework** | | | | | |
| Women (Weekday) | 977 | 195.54 | 147.84 | 0.00 | 1280.00 |
| Men (Weekday) | 977 | 44.41 | 86.46 | 0.00 | 780.00 |
| Gap (Weekday) | 977 | 151.13 | 171.55 | -610.00 | 1280.00 |
| Women (Weekend) | 977 | 175.27 | 137.21 | 0.00 | 900.00 |
| Men (Weekend) | 977 | 64.02 | 109.38 | 0.00 | 1020.00 |
| Gap (Weekend) | 977 | 111.25 | 171.72 | -670.00 | 900.00 |
| **Total time with children** | | | | | |
| Women (Weekday) | 977 | 531.49 | 268.33 | 0.00 | 1440.00 |
| Men (Weekday) | 977 | 635.42 | 251.47 | 0.00 | 1440.00 |
| Gap (Weekday) | | | | | |
| Women (Weekend) | 977 | 416.75 | 232.53 | 0.00 | 1320.00 |
| Men (Weekend) | 977 | 529.08 | 235.93 | 0.00 | 1320.00 |
| Gap (Weekend) | | | | | |
| **Engaged time with children** | | | | | |
| Women (Weekday) | 977 | 339.31 | 198.91 | 0.00 | 1080.00 |
| Men (Weekday) | | | | | |
| Gap (Weekday) | 977 | 284.93 | 232.82 | 0.00 | 1230.00 |
| Women (Weekend) | 977 | 485.91 | 282.78 | 0.00 | 1380.00 |
| Men (Weekend) | 977 | 240.07 | 209.92 | 0.00 | 1230.00 |
| Gap (Weekend) | 977 | 420.01 | 265.75 | 0.00 | 1380.00 |
| **Quality time with children** | | | | | |
| Women (Weekday) | 977 | 244.91 | 193.14 | 0.00 | 960.00 |
| Men (Weekday) | 977 | 139.23 | 231.00 | -810.00 | 1200.00 |
| Gap (Weekday) | 977 | 94.40 | 250.58 | -780.00 | 960.00 |
| Women (Weekend) | 977 | 176.67 | 282.69 | -1080.00 | 1240.00 |
| Men (Weekend) | 977 | 109.07 | 302.14 | -820.00 | 1230.00 |
| Gap (Weekend) | 977 | 246.56 | 334.41 | -1140.00 | 1320.00 |
|  | 977 | 149.52 | 332.23 | -960.00 | 1230.00 |

*Notes:* The table reports descriptive statistics for time-use variables, measured in minutes per day.

Table 13: Descriptive statistics for traditional gender norms

|  | N | Mean | SD | Min | Max |
|---|---|---|---|---|---|
| **Traditional norms index** | 3,856 | 43.34 | 23.67 | 0.00 | 100.00 |
| Traditional family roles | 3,856 | 37.46 | 33.74 | 0.00 | 100.00 |
| Child suffers if mother works (0–6) | 3,856 | 57.48 | 33.05 | 0.00 | 100.00 |
| Child suffers if mother works (7–11) | 3,856 | 48.17 | 32.94 | 0.00 | 100.00 |
| Duty to have children | 3,856 | 31.60 | 33.26 | 0.00 | 100.00 |
| Woman earns more than man | 3,856 | 38.94 | 32.60 | 0.00 | 100.00 |
| Man takes care of the home | 3,856 | 39.88 | 32.51 | 0.00 | 100.00 |
| Woman reduces aspirations | 3,856 | 49.82 | 33.38 | 0.00 | 100.00 |

*Notes:* The table reports descriptive statistics for indicators of gender attitudes. Individual questions are measured on a 0-100 agreement scale, where higher values indicate more conservative views. Composite indices are computed as averages of the relevant items.

Table 14: Descriptive statistics for traditional gender norms (Emilia-Romagna)

|  | N | Mean | SD | Min | Max |
|---|---|---|---|---|---|
| **Traditional norms index** | 1,902 | 46.70 | 23.53 | 0.00 | 100.00 |
| Traditional family roles | 1,902 | 39.59 | 33.56 | 0.00 | 100.00 |
| Child suffers if mother works (0–6) | 1,902 | 59.98 | 33.45 | 0.00 | 100.00 |
| Child suffers if mother works (7–11) | 1,902 | 51.93 | 34.37 | 0.00 | 100.00 |
| Duty to have children | 1,902 | 32.91 | 32.89 | 0.00 | 100.00 |
| Woman earns more than man | 1,902 | 42.85 | 32.35 | 0.00 | 100.00 |
| Man takes care of the home | 1,902 | 42.45 | 31.47 | 0.00 | 100.00 |
| Woman reduces aspirations | 1,902 | 57.18 | 32.54 | 0.00 | 100.00 |

*Notes:* The table reports descriptive statistics for indicators of gender attitudes. Individual questions are measured on a 0-100 agreement scale, where higher values indicate more conservative views. Composite indices are computed as averages of the relevant items.

Table 15: Descriptive statistics for traditional gender norms (Campania)

|  | N | Mean | SD | Min | Max |
|---|---|---|---|---|---|
| **Traditional norms index** | 1,954 | 40.07 | 23.36 | 0.00 | 100.00 |
| Traditional family roles | 1,954 | 35.40 | 33.79 | 0.00 | 100.00 |
| Child suffers if mother works (0–6) | 1,954 | 55.05 | 32.48 | 0.00 | 100.00 |
| Child suffers if mother works (7–11) | 1,954 | 44.50 | 31.06 | 0.00 | 100.00 |
| Duty to have children | 1,954 | 30.33 | 33.57 | 0.00 | 100.00 |
| Woman earns more than man | 1,954 | 35.14 | 32.41 | 0.00 | 100.00 |
| Man takes care of the home | 1,954 | 37.38 | 33.32 | 0.00 | 100.00 |
| Woman reduces aspirations | 1,954 | 42.66 | 32.62 | 0.00 | 100.00 |

*Notes:* The table reports descriptive statistics for indicators of gender attitudes. Individual questions are measured on a 0-100 agreement scale, where higher values indicate more conservative views. Composite indices are computed as averages of the relevant items.

## B. Variable definitions

**Paid work**

All activities that generate labor income. This includes work performed on site or remotely, time spent in main or secondary jobs, overtime, work carried out from home, and commuting time to and from the workplace. The definition follows Gimenez-Nadal and Sevilla [14].

**Unpaid work**

All activities that do not generate labor income. In this paper, unpaid work includes domestic work and childcare. It covers meal preparation, cleaning, laundry, ironing, shopping, household planning and management, small repairs, and all forms of direct childcare. The definition corresponds closely to Gimenez-Nadal and Sevilla [14].

**Leisure**

Social activities, hobbies, personal care, sleep, rest, and passive time. It includes reading, television, social media, internet use, sports, artistic activities, socializing, eating out, and other non-work activities. The concept is broader than that in Aguiar and Hurst [1] because it also includes sleep and personal care.

**Childcare**

Time spent feeding, dressing, washing, supervising, reading to children, playing with them, helping with homework, accompanying them, discussing their activities with teachers or other adults, providing health care, and other child-related tasks.

**Housework**

Time devoted to cooking, cleaning, tidying, organizing or repairing the home, laundry, shopping, planning family life, and other domestic activities [6].

**Care for other people**

Time devoted to the care of other family members who are not children.

**Total time with children**

Minutes in which at least one co-resident child is present during the activity [8].

**Engaged time with children**

Minutes in which at least one co-resident child is both present and actively involved in the activity [8].

**Quality time with children**

Enriching activities carried out with at least one co-resident child, including reading, playing, talking and listening, arts and crafts, eating together, sports, cultural activities, religious practice, physical care, and homework help for older children [18].

# C. Subsample of couples in which both partners work full-time

Table 16: Gender differences in time use (full-time couples)

| Activity | Women | Men | Diff. | t-statistic | p-value |
|---|---|---|---|---|---|
| **Paid work** | | | | | |
| Weekday | 362,6 | 432,7 | -70,1 | -4,57 | 0,000 |
| Weekend | 58,4 | 110,1 | -51,7 | -4,29 | 0,000 |
| **Unpaid work** | | | | | |
| Weekday | 219,5 | 128,7 | 90,8 | 10,16 | 0,000 |
| Weekend | 288,3 | 206,2 | 82,2 | 7,14 | 0,000 |
| **Total work** | | | | | |
| Weekday | 582,1 | 561,3 | 20,7 | 1,37 | 0,172 |
| Weekend | 346,7 | 316,2 | 30,4 | 2,04 | 0,041 |
| **Leisure** | | | | | |
| Weekday | 795,9 | 817,0 | -21,1 | -1,45 | 0,148 |
| Weekend | 986,4 | 1025,2 | -38,7 | -2,53 | 0,012 |
| **Leisure with children** | | | | | |
| Weekday | 185,5 | 169,8 | 15,7 | 1,36 | 0,173 |
| Weekend | 328,6 | 312,3 | 16,3 | 1,16 | 0,247 |
| **Leisure without children** | | | | | |
| Weekday | 610,4 | 647,2 | -36,8 | -3,51 | 0,000 |
| Weekend | 657,9 | 712,9 | -55,1 | -4,75 | 0,000 |
| **Childcare** | | | | | |
| Weekday | 119,5 | 89,9 | 29,6 | 4,53 | 0,000 |
| Weekend | 146,3 | 126,6 | 19,7 | 2,06 | 0,040 |
| **Housework** | | | | | |
| Weekday | 100,0 | 38,8 | 61,2 | 11,13 | 0,000 |
| Weekend | 142,0 | 78,9 | 63,1 | 8,82 | 0,000 |
| **Care for other people** | | | | | |
| Weekday | 0,0 | 0,0 | 0,0 | | |
| Weekend | 0,0 | 0,6 | -0,6 | -1,67 | 0,095 |
| **Total time with children** | | | | | |
| Weekday | 388,1 | 296,8 | 91,3 | 6,22 | 0,000 |
| Weekend | 630,6 | 545,3 | 85,3 | 5,43 | 0,000 |
| **Engaged time with children** | | | | | |
| Weekday | 329,8 | 264,2 | 65,6 | 4,87 | 0,000 |
| Weekend | 552,3 | 493,9 | 58,3 | 3,71 | 0,000 |
| **Quality time with children** | | | | | |
| Weekday | 211,0 | 164,9 | 46,1 | 5,10 | 0,000 |
| Weekend | 310,9 | 258,4 | 52,5 | 4,56 | 0,000 |
| N | 512 | 512 | 512 | | |

*Notes:* The table reports average daily time devoted to each activity, measured in minutes. The difference is computed as women minus men. t-statistics and p-values come from tests of equality of means across the two groups.

Table 17: Gender differences in time use (Emilia-Romagna, full-time couples)

| Activity | Women | Men | Diff. | t-statistic | p-value |
|---|---|---|---|---|---|
| **Paid work** | | | | | |
| Weekday | 371,1 | 430,3 | -59,2 | -3,30 | 0,001 |
| Weekend | 43,3 | 90,4 | -47,1 | -3,73 | 0,000 |
| **Unpaid work** | | | | | |
| Weekday | 189,3 | 120,7 | 68,5 | 7,65 | 0,000 |
| Weekend | 273,4 | 204,2 | 69,1 | 5,47 | 0,000 |
| **Total work** | | | | | |
| Weekday | 560,4 | 551,0 | 9,3 | 0,52 | 0,603 |
| Weekend | 316,7 | 294,6 | 22,1 | 1,37 | 0,172 |
| **Leisure** | | | | | |
| Weekday | 820,7 | 830,3 | -9,6 | -0,55 | 0,583 |
| Weekend | 1010,8 | 1039,3 | -28,5 | -1,66 | 0,097 |
| **Leisure with children** | | | | | |
| Weekday | 201,0 | 177,1 | 23,9 | 1,73 | 0,084 |
| Weekend | 335,5 | 324,2 | 11,3 | 0,70 | 0,484 |
| **Leisure without children** | | | | | |
| Weekday | 619,7 | 653,2 | -33,5 | -2,66 | 0,008 |
| Weekend | 675,3 | 715,1 | -39,8 | -3,01 | 0,003 |
| **Childcare** | | | | | |
| Weekday | 98,2 | 82,5 | 15,7 | 2,49 | 0,013 |
| Weekend | 128,4 | 122,6 | 5,8 | 0,57 | 0,569 |
| **Housework** | | | | | |
| Weekday | 91,1 | 38,2 | 52,8 | 8,67 | 0,000 |
| Weekend | 145,0 | 81,2 | 63,8 | 7,72 | 0,000 |
| **Care for other people** | | | | | |
| Weekday | 0,0 | 0,0 | 0,0 | | |
| Weekend | 0,0 | 0,5 | -0,5 | -1,34 | 0,180 |
| **Total time with children** | | | | | |
| Weekday | 378,5 | 299,3 | 79,2 | 4,70 | 0,000 |
| Weekend | 629,0 | 563,9 | 65,1 | 3,71 | 0,000 |
| **Engaged time with children** | | | | | |
| Weekday | 326,6 | 268,8 | 57,8 | 3,74 | 0,000 |
| Weekend | 552,7 | 509,8 | 42,9 | 2,41 | 0,016 |
| **Quality time with children** | | | | | |
| Weekday | 194,5 | 162,4 | 32,1 | 3,28 | 0,001 |
| Weekend | 298,3 | 252,1 | 46,2 | 3,66 | 0,000 |
| N | 392 | 392 | 392 | | |

*Notes:* The table reports average daily time devoted to each activity, measured in minutes. The difference is computed as women minus men. t-statistics and p-values come from tests of equality of means across the two groups.

Table 18: Gender differences in time use (Campania, full-time couples)

| Activity | Women | Men | Diff. | t-statistic | p-value |
|---|---|---|---|---|---|
| **Paid work** | | | | | |
| Weekday | 334,7 | 440,3 | -105,7 | -3,63 | 0,000 |
| Weekend | 107,6 | 174,5 | -66,9 | -2,26 | 0,025 |
| **Unpaid work** | | | | | |
| Weekday | 318,2 | 154,6 | 163,7 | 7,41 | 0,000 |
| Weekend | 337,2 | 212,4 | 124,8 | 4,78 | 0,000 |
| **Total work** | | | | | |
| Weekday | 652,9 | 594,9 | 58,0 | 2,20 | 0,029 |
| Weekend | 444,8 | 386,9 | 57,8 | 1,73 | 0,085 |
| **Leisure** | | | | | |
| Weekday | 714,8 | 773,4 | -58,6 | -2,57 | 0,011 |
| Weekend | 907,0 | 979,1 | -72,1 | -2,23 | 0,027 |
| **Leisure with children** | | | | | |
| Weekday | 134,8 | 145,8 | -11,0 | -0,59 | 0,559 |
| Weekend | 306,1 | 273,4 | 32,7 | 1,15 | 0,252 |
| **Leisure without children** | | | | | |
| Weekday | 580,1 | 627,7 | -47,6 | -2,80 | 0,006 |
| Weekend | 600,9 | 705,7 | -104,8 | -4,47 | 0,000 |
| **Childcare** | | | | | |
| Weekday | 189,0 | 113,9 | 75,1 | 4,36 | 0,000 |
| Weekend | 204,9 | 140,0 | 64,9 | 2,87 | 0,005 |
| **Housework** | | | | | |
| Weekday | 129,2 | 40,7 | 88,6 | 7,33 | 0,000 |
| Weekend | 132,2 | 71,4 | 60,8 | 4,28 | 0,000 |
| **Care for other people** | | | | | |
| Weekday | 0,0 | 0,0 | 0,0 | | |
| Weekend | 0,0 | 1,0 | -1,0 | -1,00 | 0,318 |
| **Total time with children** | | | | | |
| Weekday | 419,8 | 288,8 | 130,9 | 4,39 | 0,000 |
| Weekend | 636,0 | 484,7 | 151,3 | 4,41 | 0,000 |
| **Engaged time with children** | | | | | |
| Weekday | 340,2 | 248,9 | 91,2 | 3,33 | 0,001 |
| Weekend | 550,8 | 442,2 | 108,6 | 3,27 | 0,001 |
| **Quality time with children** | | | | | |
| Weekday | 264,9 | 173,1 | 91,8 | 4,38 | 0,000 |
| Weekend | 352,1 | 279,0 | 73,1 | 2,77 | 0,006 |
| N | 120 | 120 | 120 | | |

*Notes:* The table reports average daily time devoted to each activity, measured in minutes. The difference is computed as women minus men. t-statistics and p-values come from tests of equality of means across the two groups.

Table 19: Gender differences in attitudes toward gender norms (full-time couples)

| Indicator | Women | Men | Diff. | t-statistic | p-value |
|---|---|---|---|---|---|
| **Traditional norms index** | 40,9 | 42,7 | -1,8 | -1,35 | 0,178 |
| Traditional family roles | 29,5 | 33,7 | -4,1 | -2,10 | 0,036 |
| Child suffers if mother works (0–6) | 56,3 | 57,9 | -1,6 | -0,76 | 0,445 |
| Child suffers if mother works (7–11) | 49,6 | 50,7 | -1,2 | -0,54 | 0,587 |
| Duty to have children | 25,2 | 29,7 | -4,5 | -2,36 | 0,019 |
| Woman earns more than man | 38,0 | 35,5 | 2,6 | 1,32 | 0,188 |
| Man takes care of the home | 36,1 | 38,8 | -2,6 | -1,37 | 0,171 |
| Woman reduces aspirations | 51,6 | 52,9 | -1,2 | -0,59 | 0,557 |
| N | 512 | 512 | 512 | | |

*Notes:* The table reports mean values for indicators of gender attitudes. Individual questions are measured on a 0-100 agreement scale, where higher values indicate more conservative views. Composite indices are computed as averages of the relevant items. The difference is computed as women minus men. t-statistics and p-values come from tests of equality of means across the two groups.

Table 20: Gender differences in attitudes toward gender norms (Emilia-Romagna, full-time couples)

| Indicator | Women | Men | Diff. | t-statistic | p-value |
|---|---|---|---|---|---|
| **Traditional norms index** | 42,9 | 44,7 | -1,8 | -1,21 | 0,227 |
| Traditional family roles | 31,3 | 35,5 | -4,2 | -1,88 | 0,061 |
| Child suffers if mother works (0–6) | 58,8 | 60,7 | -2,0 | -0,83 | 0,408 |
| Child suffers if mother works (7–11) | 51,7 | 53,3 | -1,5 | -0,62 | 0,533 |
| Duty to have children | 24,7 | 28,8 | -4,2 | -1,96 | 0,050 |
| Woman earns more than man | 39,9 | 37,2 | 2,7 | 1,22 | 0,221 |
| Man takes care of the home | 37,6 | 40,0 | -2,4 | -1,09 | 0,275 |
| Woman reduces aspirations | 56,6 | 57,5 | -1,0 | -0,40 | 0,687 |
| N | 392 | 392 | 392 | | |

*Notes:* The table reports mean values for indicators of gender attitudes. Individual questions are measured on a 0-100 agreement scale, where higher values indicate more conservative views. Composite indices are computed as averages of the relevant items. The difference is computed as women minus men. t-statistics and p-values come from tests of equality of means across the two groups.

Table 21: Gender differences in attitudes toward gender norms (Campania, full-time couples)

| Indicator | Women | Men | Diff. | t-statistic | p-value |
|---|---|---|---|---|---|
| **Traditional norms index** | 34,3 | 36,2 | -1,9 | -0,64 | 0,521 |
| Traditional family roles | 23,8 | 27,8 | -4,0 | -0,97 | 0,331 |
| Child suffers if mother works (0–6) | 48,0 | 48,5 | -0,4 | -0,10 | 0,919 |
| Child suffers if mother works (7–11) | 42,5 | 42,3 | 0,1 | 0,03 | 0,978 |
| Duty to have children | 27,0 | 32,7 | -5,7 | -1,30 | 0,193 |
| Woman earns more than man | 31,9 | 29,8 | 2,1 | 0,52 | 0,602 |
| Man takes care of the home | 31,4 | 34,8 | -3,5 | -0,85 | 0,394 |
| Woman reduces aspirations | 35,4 | 37,6 | -2,2 | -0,54 | 0,587 |
| N | 120 | 120 | 120 | | |

*Notes:* The table reports mean values for indicators of gender attitudes. Individual questions are measured on a 0-100 agreement scale, where higher values indicate more conservative views. Composite indices are computed as averages of the relevant items. The difference is computed as women minus men. t-statistics and p-values come from tests of equality of means across the two groups.